\documentclass[10pt,conference]{IEEEtran}
\usepackage{cite}
\usepackage{amsmath,amssymb,amsfonts}
\usepackage{algorithmic}
\usepackage{graphicx}
\usepackage{textcomp}
\usepackage{xcolor}
\usepackage[hyphens]{url}
\usepackage{fancyhdr}
\usepackage{hyperref}

\newcommand{\hpcayear}{2025}

\newcommand{\hpcasubmissionnumber}{1951}
\title{DL-PIM: Improving Data Locality in Processing-in-Memory Systems}
\def\hpcacameraready{} % Uncomment to build camera-ready version

\newcommand\hpcaauthors{Parker Hao Tian$^ 1$$^ 2$, Zahra Yousefijamarani$^ 2$, Alaa Alameldeen$^ 2$}
\newcommand\hpcaaffiliation{Amazon$^ 1$, Simon Fraser University$^ 2$}
\newcommand\hpcaemail{ }

\author{
  \ifdefined\hpcacameraready
    \IEEEauthorblockN{\hpcaauthors{}}
      \IEEEauthorblockA{
        \hpcaaffiliation{} \\
        \hpcaemail{}
      }
  \else
    \IEEEauthorblockN{\normalsize{HPCA \hpcayear{} Submission
      \textbf{\#\hpcasubmissionnumber{}}} \\
      \IEEEauthorblockA{
        Confidential Draft \\
        Do NOT Distribute!!
      }
    }
  \fi 
}

\usepackage{multirow}
\usepackage{tabularray}

\begin{document}
\maketitle

\newcommand{\hpcaheight}{0mm}
\ifdefined\eaopen
\renewcommand{\hpcaheight}{12mm}
\fi

%%%%%%%%%%%%%%%%%%%%%%%%%%%%%%%%%%%%%%%%
%%%%%%%% -- PAPER CONTENT STARTS -- %%%%%%%%%

\begin{abstract}
Processing-in-Memory (PIM) architectures aim to reduce data transfer costs between processors and memory by integrating processing units within memory layers. Prior PIM architectures have shown potential to improve energy efficiency and performance. However, such advantages rely on data proximity to the processing unit(s) performing computations. Data movement overheads can degrade PIM's performance and energy efficiency due to the need to move data between a processing unit and a distant memory location. 
%they face challenges due to the overhead of transferring data from remote memory locations to processing units inside memory for computation.

In this paper, we demonstrate that a large fraction of PIM's latency per memory request is attributed to data transfers and queuing delays from remote memory accesses. To improve PIM's data locality, we propose DL-PIM, a novel architecture that dynamically detects the overhead of data movement, and proactively moves data to a reserved area in the local memory of the requesting processing unit. DL-PIM uses a distributed address-indirection hardware lookup table to redirect traffic to the current data location. We propose DL-PIM implementations on two 3D stacked memories: Hybrid Memory Cube (HMC) and High Bandwidth Memory (HBM). While some workloads benefit from DL-PIM, others are negatively impacted by the additional latency due to indirection accesses. Therefore, we propose an adaptive mechanism that assesses the cost and benefit of indirection and dynamically enables or disables it to prevent degrading workloads that suffer from indirection. Overall, DL-PIM reduces the average memory latency per request by 54\% in HMC and 50\% in HBM which resulted in performance improvement of 15\% for workloads with substantial data reuse in HMC and 5\% in HBM. For all representative workloads, DL-PIM achieved a 6\% speedup in HMC and a 3\% speedup in HBM, showing that DL-PIM enhances data locality and overall system performance.

\end{abstract}
\section{Introduction}

Many sectors in today's global economy are highly dependent on applications such as machine learning, data analytics, graph
analytics, transaction processing, and scientific high-performance computing. Such Big Data applications have exponentially growing data needs, leading to an increasing memory footprint and higher dependence on memory speed and bandwidth
%Modern workloads ranging from machine learning and artificial intelligence to genome analysis and scientific high-performance computing, are experiencing an exponential escalation in data demands. Consequently, the necessity for rapid data analysis is intensifying. This trend shows that main memory is becoming more important, but it's also becoming a critical bottleneck in system performance 
\cite{mutlu2021intelligent, pimworkloaddriven, bigdata-memory}. 
In compute-centric systems, moving data between memory and processors consumes a significant amount of energy and causes performance loss, especially for workloads that have poor temporal and spatial locality. To reduce the data movement overhead, processing-in-memory (PIM) architectures have been proposed. In PIM systems, computation is brought 
%A major reason for the bottleneck of main memory is the high energy and latency associated with data movement between processor and main memory \cite{INVITED}. 
%To address this bottleneck, one solution is to bring computation 
closer to the data, instead of moving data across the system to distant compute units. Recent PIM proposals (also known as near-data processing) \cite{ambit,prime,scalablepimforgraphprocessing, simdram, akin2015data} avoided such unnecessary data movement. This approach has been adopted across various applications   including graph processing \cite{graphh,scalablepimforgraphprocessing,graphq,graphp}, neural networks \cite{tetrisnn}, sparse matrix-vector multiplication \cite{spacea}, and bioinformatics \cite{kim2018grim}.
%have been widely adopted PIM to accelerate computational tasks.

PIM systems improve both performance and energy efficiency. For example, using 64-bit fully functional ARM-like PIM cores saves 50.9\% of energy energy consumption and improves performance by up to 57.2\% in the TensorFlow Lite workload \cite{pimworkloaddriven}.  

\textbf{Limitations of the state of the art.} While PIM has shown to be promising for many applications, it faces some challenges and has been limited by how the memory architecture is implemented \cite{oliveira2021damov}. For instance, Micron's Hybrid Memory Cube (HMC), a stacked memory architecture widely explored for PIM, divides the memory system into separate vaults \cite{micron}. When a PIM processing unit requires access to data from another vault, significant overhead is incurred as data is transferred between the home and requesting vaults, resulting in data transfer latency or network latency.

Another significant factor contributing to the overall latency overhead in PIM is queuing delay. Accesses to various vaults may be unbalanced, with memory locations in a vault accessible by multiple vaults simultaneously. However, since each vault can only serve one location per cycle, memory requests must be queued, resulting in considerable delays. 

\begin{figure}[t]
% \vspace{-5pt}
\centering
\includegraphics[width=0.5\textwidth]{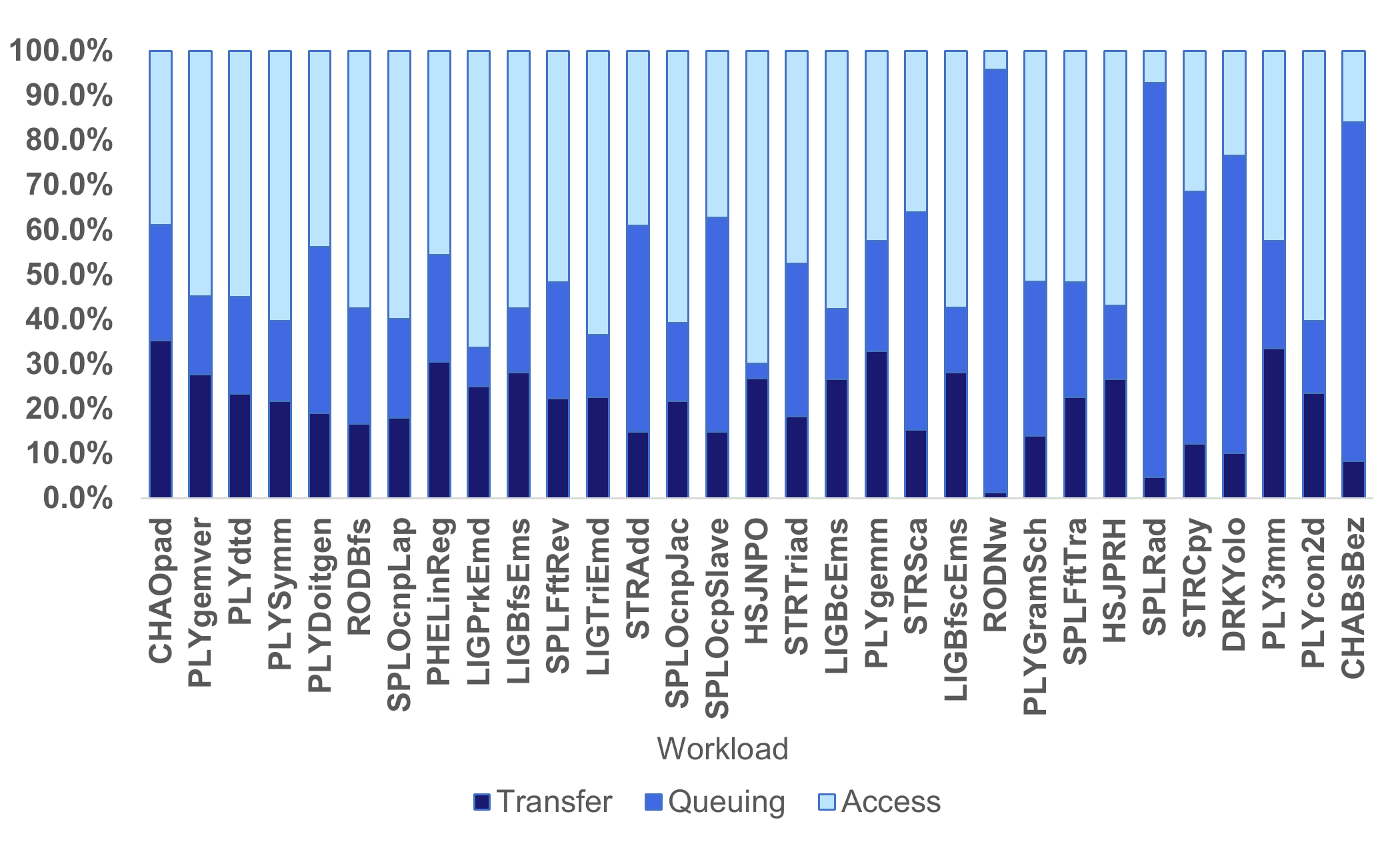}
\vspace{-15pt}
\caption{Breakdown of Memory Latency into data transfer latency, queuing delay and array access latency with HMC memory.}
\label{fig:memory_latency_breakdown}
\end{figure}

\begin{figure}[t]
\vspace{-5pt}
\centering
\includegraphics[width=0.5\textwidth]{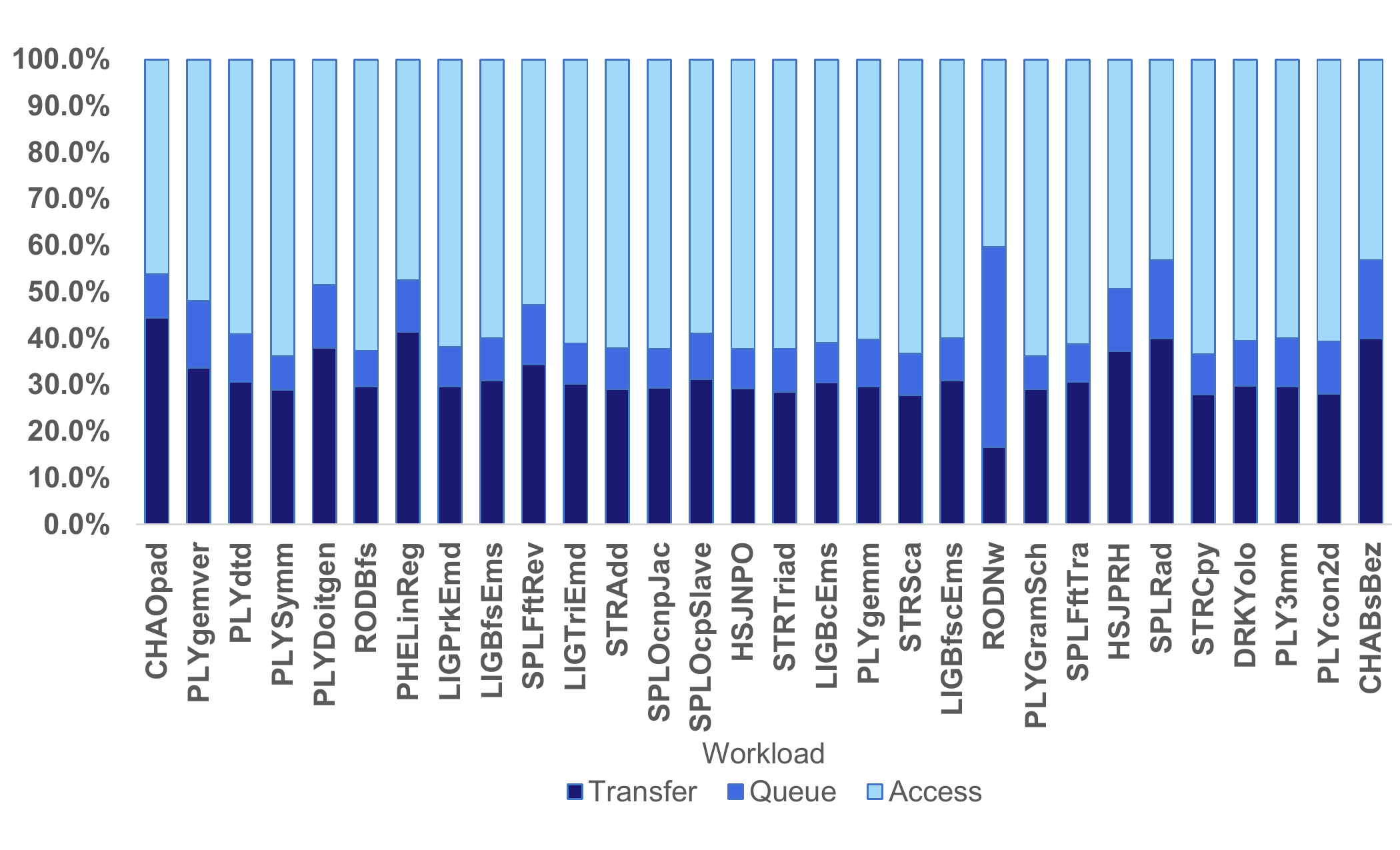}
\vspace{-20pt}
\caption{Breakdown of Memory Latency with HBM.}
\label{fig:memory_latency_breakdown_HBM}
\vspace{-5pt}
\end{figure}

Through simulations of various common workloads using the DAMOV framework \cite{oliveira2021damov}, we observed that in HMC with $6\times6$ network, approximately 53\% of the latency per memory request is attributed to data transfers and queuing delays from remote memory accesses, as illustrated in Figure~\ref{fig:memory_latency_breakdown}. For high bandwidth memory(HBM) with $4\times2$ network, which is typical of HBM architectures, this ratio is about 43\% as shown in Figure~\ref{fig:memory_latency_breakdown_HBM}. Both figures show breakdown of memory latency into data transfer (network) latency, queuing delay, and array access latency. The higher queuing delay in HMC compared to HBM is due to the larger network in HMC, which requires data packets to traverse multiple hops and deal with more complex interconnections. Additionally, some workloads exhibit significantly higher queuing delays than others. To investigate this further, we examined the coefficient of variation (CoV) across different vaults to understand the underlying reasons. 

\begin{figure}[t]
\centering
\includegraphics[width=0.5\textwidth]{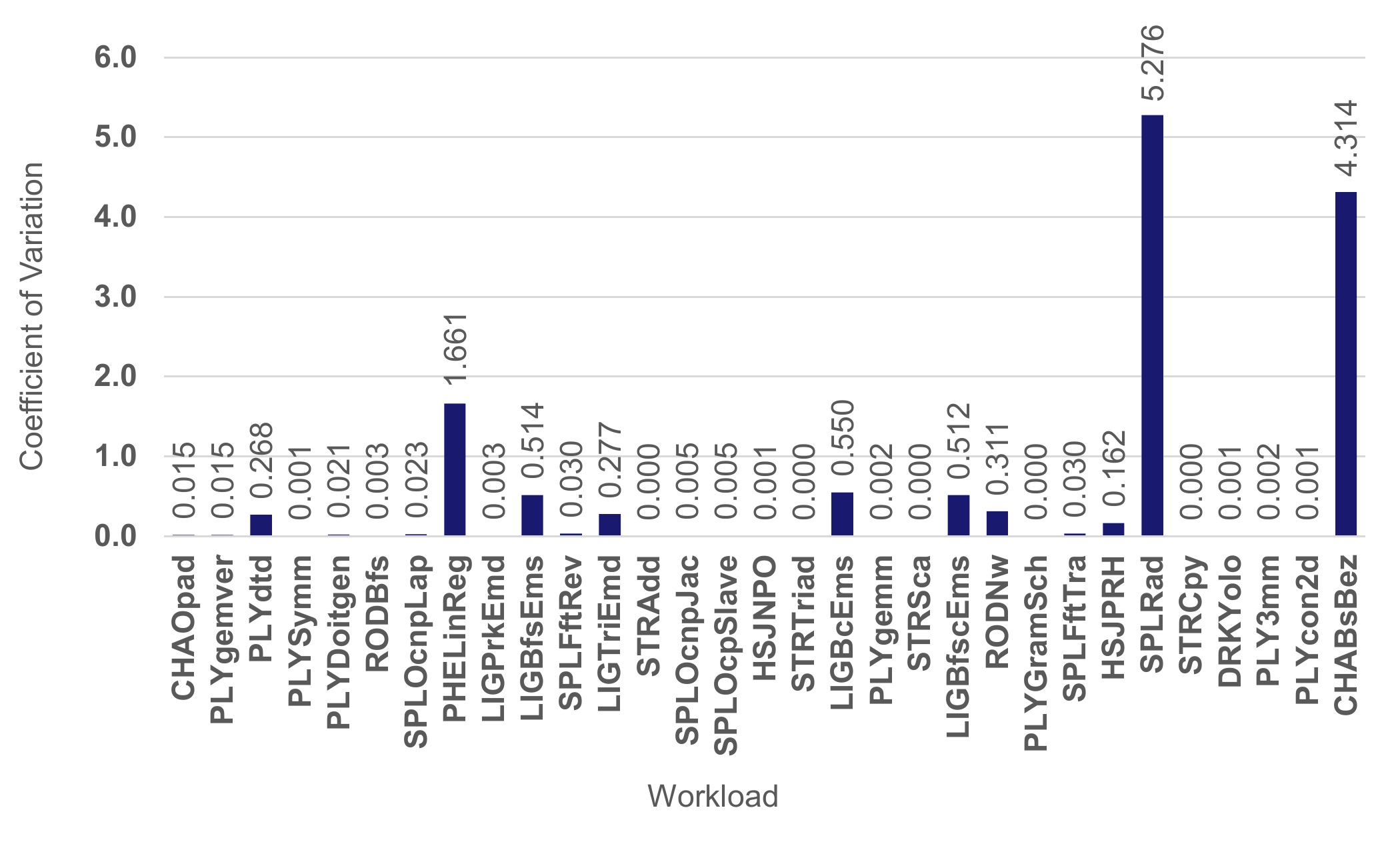}
\vspace{-15pt}
\caption{Coefficient of variation (CoV) for memory request distribution across workloads with HMC memory.}
\label{fig:cov-baseline}
\vspace{-5pt}
\end{figure}

Figure~\ref{fig:cov-baseline} illustrates the CoV of this distribution within a $6\times6$ HMC system. A high CoV suggests that certain vaults experience substantially greater demand compared to others. While most workloads exhibit balanced access distribution across vaults (low CoV), some demonstrate highly imbalanced distributions, leading to considerable additional queuing delays at high-demand vaults. This phenomenon is highlighted in Figure~\ref{fig:memory_latency_breakdown}, where workloads with high CoV values can attribute 70-80\% of their memory latency to queuing. We discuss these workloads further in Section~\ref{sec_evaluation}.

\begin{figure}[t]
\vspace{-5pt}
\centering
\includegraphics[width=0.5\textwidth]{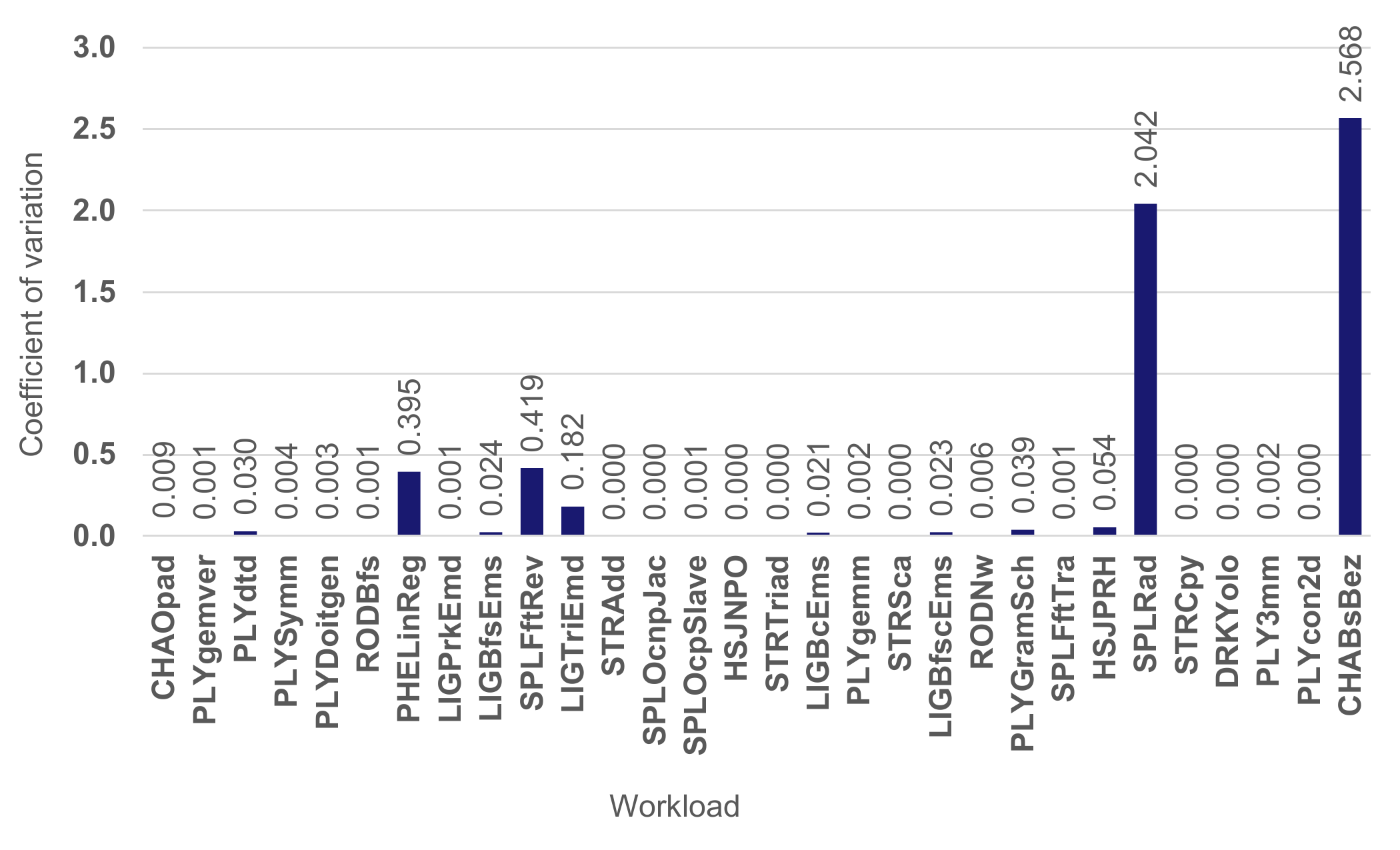}
\vspace{-15pt}
\caption{CoV for memory request distribution across workloads with HBM.}
\label{fig:cov-baseline_hbm}
\vspace{-5pt}
\end{figure}

We also investigated CoV for memory request distribution in HBM PIM systems, as shown in Figure~\ref{fig:cov-baseline_hbm}. Our findings reveal that while some workloads in the HBM PIM system exhibit higher CoV values, indicating a noticeable imbalance in memory request distribution similar to the patterns observed in Figure\ref{fig:cov-baseline} for HMC systems, the overall CoV values in HBM are generally lower. The lower CoV in HBM indicates a more even distribution of memory requests across different vaults compared to HMC. As a result, HBM systems experience less severe queuing delays due to imbalances in demand as shown in Figure~\ref{fig:memory_latency_breakdown_HBM}.

\textbf{Our Proposal: DL-PIM.} To reduce data transfer and queuing delays, an efficient PIM architecture should place data near the processing element where it is mostly accessed. However, optimal placement is impractical since it requires prior knowledge of which processing elements need to access different program data locations. In this paper, we attempt to reduce data transfer and queuing delays across memory modules ("vaults" for HMC) in a PIM system using a hardware indirection-based mechanism. Our architecture, Data Locality-based PIM (DL-PIM) reserves a small memory area at each vault, and dynamically moves a memory block into that reserved space if the benefit of movement outweighs the cost. We maintain a distributed hardware \textit{subscription table} where each vault tracks local blocks that moved to remote vaults, and remote blocks that moved to the current vault. By examining runtime information across memory requests, DL-PIM can detect when subscriptions help or hurt performance of various workloads, and therefore dynamically turn subscriptions on or off.

\textbf{Contributions.} In this paper, we make the following contributions: 
\begin{itemize}
    \item We demonstrate the high overhead caused by data movement \textit{within a PIM system}, unlike other works focusing on processor-memory data movement.
    \item We propose an \textit{always-subscribe} architecture that dynamically moves data to requesting memory modules, which helps some workloads while negatively impacting others. 
    \item We propose an \textit{adaptive} mechanism that turns on subscription only when it benefits a workload's performance.
    \item We show that DL-PIM implemented in HMC, improves average performance by 6\% for all workloads, and by 15\% (and up to more than 2X) for workloads that have non-negligible data reuse. This is caused by a 54\% reduction in average memory latency per requests for these workloads.
    \item We show that DL-PIM implemented in HBM, reduced average memory latency by 50\%, improved performance by 3\% for all workloads and 5\% for workloads with high data reuse.  
\end{itemize}

\section{Background}
\label{sec_background}

\subsection{Memory Wall}
With the exponential improvement in CPU performance over the past few decades, memory has become the primary performance bottleneck for many important applications. Modern processors use a multi-level cache hierarchy that exploits spatial and temporal locality to reduce average memory access time. When data is reused in the cache hierarchy, the cost of data transfer to/from memory is amortized across many cache hits. As main memory is much slower than caches, cache misses incur significant performance overhead. Unfortunately, many workloads exhibit poor locality where cache data is reused infrequently, which requires high time and energy overheads to continuously transfer data from memory (where it resides) to processors (where computations are performed). This \textit{memory wall}~\cite{reflectiononmemwall} affects processor-centric systems and degrades the performance of data-intensive workloads, e.g., database systems, machine learning and graph processing.

\subsection{Processing in Memory}
 
To mitigate the performance and energy overheads associated with the continuous transfer of data between processors and memory, memory-centric systems have been extensively investigated over the past few decades, gaining significant traction in both academia and industry. Processing-in-Memory (PIM) architectures, which integrate computational capabilities directly into the memory system, have emerged as a promising solution \cite{missingthememwall}. PIM exploits the high internal bandwidth of memory, eliminating the need to transfer data to and from the CPU, thus conserving both time and energy.

PIM hardware implementations can be classified into two categories: processing near memory and processing using memory \cite{pimworkloaddriven}. Both categories address the data movement bottleneck in different ways. Processing near memory architectures place processing units near memory, e.g., in the logic layer of a stacked memory technology, to provide high bandwidth and low latency communication between the processing unit and the memory. Processing using memory architectures use the architecture and properties of memory cells to allow for data operations without the involvement of a CPU or accelerator \cite{pimworkloaddriven}. Although there are many proposals for processing using memory architectures (~\cite{computecaches, neuralcache, dualitycache, ambit, seshadri2016buddyram, seshadri2016processing, rowclone, simdram, drstrange, flashcosmos, energyefficientvlsi, lisa, fastbulkbitwiseandandor, drisa, gatherscatterdram, angizi2019accelerating, pinatubo, pimalogic, cmppim, aligns, levy_logic_2014, magic, isaac, imply, implyprinciplesandmethodologies, plim, revamp, memristorbasedcomputation, fastboolean, memristorforcomputing, memristivedevices}), they present significant complexity due to the need to modify standard memory interfaces and internal hardware. On the other hand, with the development of 3D-stacked memories like HMC~\cite{micron} and HBM~\cite{sohn20161}, processing near memory architectures (~\cite{fpgabasednmp, pimenabledinsts, impica, upmemtruepimaccelerator, transpimlib, axdimm, industrialproduct, a20nm, googleworkload, conda, lazypim, napel, chameleon, jafar, prime, nda, practicalneardataprocessing, hrl, biscuit, guo20143d, continuousrenahead, enhancedmemorycontroller, neurocube, bssync, pimenabledinstructions, genstore}) have been gaining traction recently.

\subsection{Hybrid Memory Cube}

HMC utilizes the 3D-stacked through-silicon-via technology that stacks dies together. HMC divides the memory system into different \textit{vaults}, each including one logical die and multiple memory dies. The logical die acts as a memory controller (called "vault controller"). A processing near memory architecture that places the processing units in the logical die benefits from high bandwidth and low latency accesses to memory. When reading from or writing to HMC from another vault or from another component, a different protocol is used. Instead of the traditional channel-based communication protocol, HMC utilizes a packet-based communication protocol. Each packet, i.e., \textit{FLIT}, is 128 bits (16B) in size. HMC supports 16B, 32B, 64B or 128B byte memory blocks. Given that HMC would also require one FLIT to store the operation information in a header and a tail, each data access may require between 2 and 9 FLITs.

When communicating with external components, e.g., processors, HMC requires communication \textit{links} with specifications like output buffers and input buffers, However, HMC does not define the required communication technology for internal communications and only uses crossbar switches as an example in the specification document \cite{micron}. In this paper, we assume that the vaults are connected in a crossbar switch network, with each vault only having input buffers of size 16 entries. 

\begin{figure}[t]
\vspace{-8pt}
\centering
\includegraphics[width=0.45\textwidth]{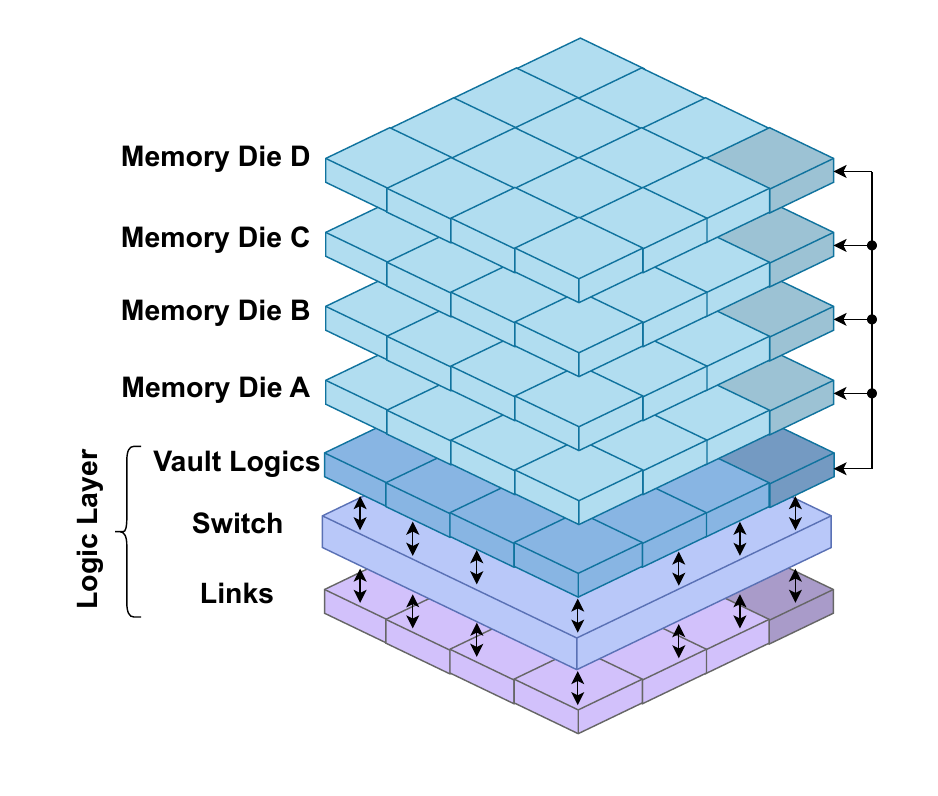}
%\vspace{-15pt}
\caption{Representation of an HMC system with 16 vaults in each layer. Each set of partitions is connected to its corresponding vault logic, illustrated with different colors in the image.}
\label{fig:HMC-16-vaults}
\vspace{-8pt}
\end{figure}

Although HMC can be used to implement a processing near memory architecture, it still faces challenging design choices. Due to the capacity limitations of each memory die, each HMC vault can have only limited capacity. As a result, an HMC system would require multiple vaults interconnected with each other (usually via a crossbar switch network) to provide a larger capacity. A 16-vault HMC system is illustrated in Figure~\ref{fig:HMC-16-vaults}. In such system, each vault acts like a router that is interconnected with the neighbouring vaults, and forwards non-local requests to the vault that is closest to its destination. Although with more vaults there are more logical dies that we could increase computational parallelism, they introduce additional complexity due to requiring remote accesses to data from other vaults. While each logic die has high bandwidth and low latency to the memory dies within its own vault, accesses to other vaults incur high data transfer and queuing delays, degrading performance and energy efficiency.

\noindent\textbf{Queuing Delay}. As HMC is a packet-based memory system, each vault can serve one memory request at a time. When multiple vaults need to access memory locations within a vault, these requests need to be queued. HMC has I/O buffers that handle this scenario, and packets arrived have to be queued in these buffers until they reach the head of the queue.

\noindent\textbf{Non-Local Access Overhead}. We use the DAMOV \cite{oliveira2021damov} framework's default $6\times6$ network to demonstrate the network data transfer and queuing overhead across workloads. Figure~\ref{fig:memory_latency_breakdown} shows that data transfer and queuing latencies account for 53\% of memory access latency on average. To address the overhead caused by non-local memory accesses, we propose a mechanism that "attracts" data to the vault where it is likely to get most accesses. Our proposal (described in the next section) attempts to reduce communication (data transfer) overhead and queuing delay. In the remainder of the paper, we call such transfer a \textit{subscription}. 

\subsection{High Bandwidth Memory}
\begin{figure}[t]
%\vspace{-10pt}
\centering
\includegraphics[width=0.45\textwidth]{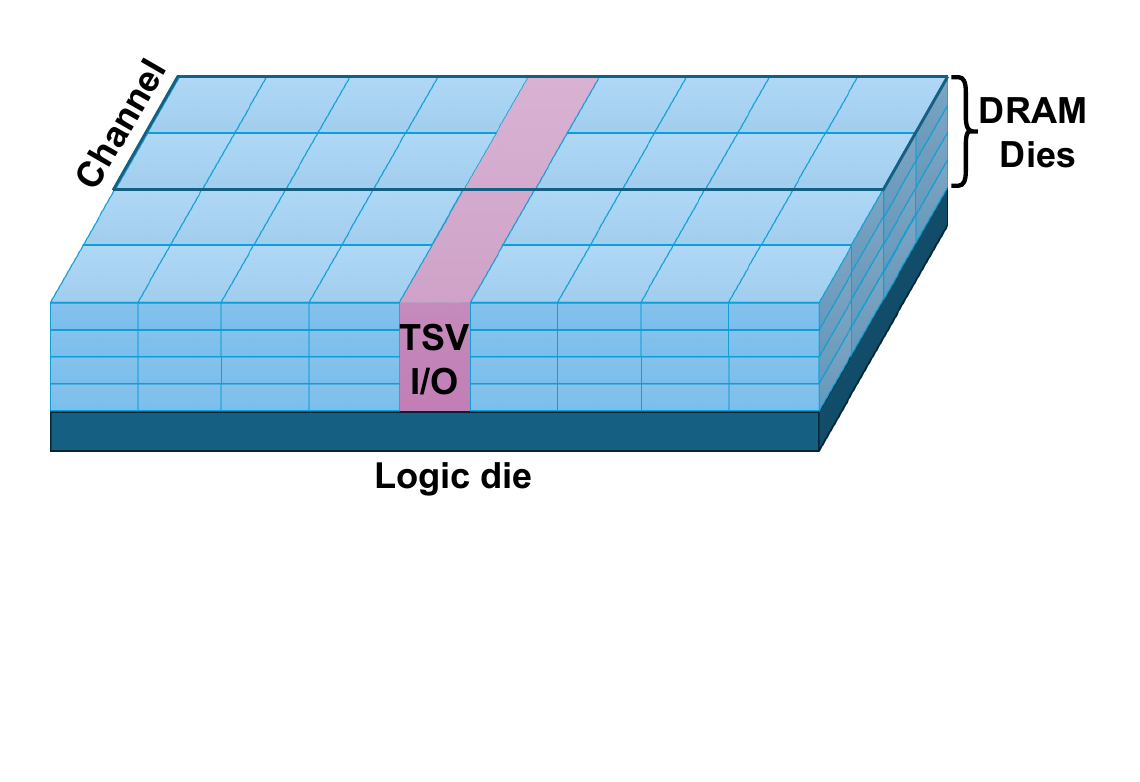}
\vspace{-60pt}
\caption{Representation of an HBM stacked DRAM architecture with 4 DRAM dies and a logic die at the bottom. Each DRAM die consisting two channels and 16 banks. }
\vspace{-5pt}
\label{fig:HBM}
\end{figure}

HBM is a memory architecture designed to deliver high bandwidth and large capacity. As shown in Figure~\ref{fig:HBM}, it consists of multiple core DRAM dies stacked vertically atop a base logic die, which manages memory operations. Each core DRAM die is divided into two channels, with each channel containing four bank groups, and each bank group consisting of four banks \cite{oh202022}. The channels operate independently with their own address and data TSVs (through-silicon vias) for point-to-point (P2P) connections, ensuring isolated operations and reducing interference. Despite the independent data pathways, the power and ground planes are shared across channels, maintaining a consistent power supply and simplifying the design \cite{lee20141}. 

The base logic die of HBM consists of several key components. The PHY handles the main interface between the HBM DRAM and the memory controller in the host. The central area of the logic die is dedicated to TSVs, which deliver signals, power, and ground to the stacked core dies. Additionally, there is a designated area for test logic and DA ports \cite{jun2017hbm}.

\section{System Overview}

Our proposal seeks to address the issues discussed in the previous section by implementing an architecture that not only dynamically "attracts" data to the vault (or channel in HBM) where it is likely to get most access, but also balances the load between vaults such that the queuing delay would be decreased. In the following discussion, we use "vault" for HMC-based PIM, but the same design can be used for channels in HBM. 

\subsection{Hardware Structures}
\begin{figure}[t]
\vspace{-10pt}
\centering
\includegraphics[width=0.45\textwidth]{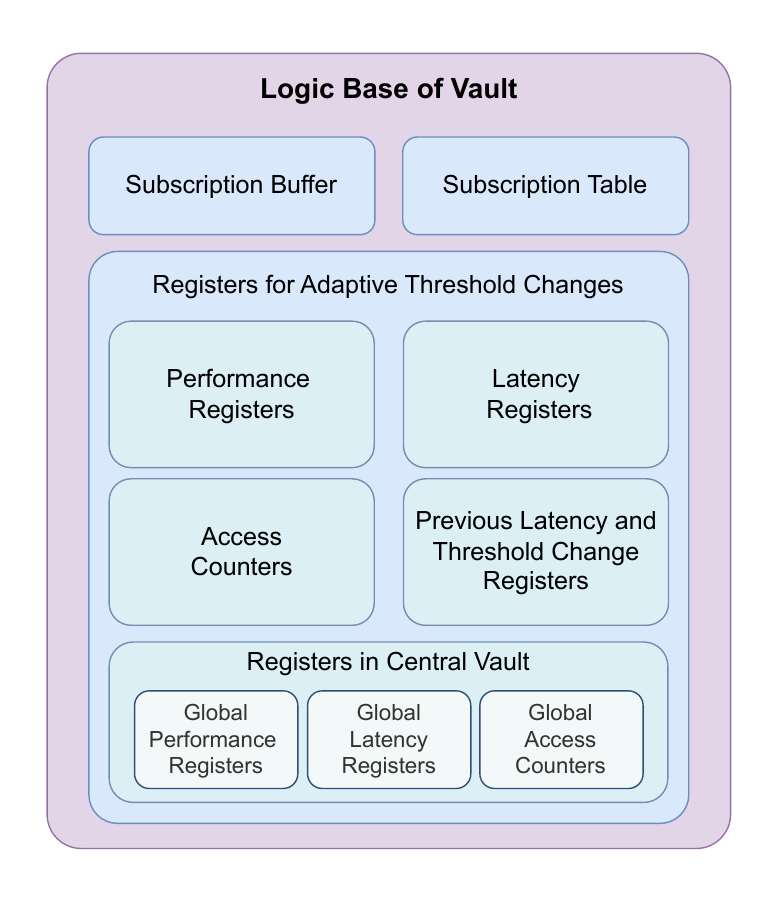}
\vspace{-5pt}
\caption{DL-PIM hardware structures added to the logical base of each vault.}
\vspace{-6pt}
\label{fig:hardware_structures}
\end{figure}

The hardware structures of our proposed system include several key components (as shown in Figure~\ref{fig:hardware_structures}) designed to facilitate efficient data management and reduce overhead:

\begin{itemize}
% \item A count table
\item \textbf{Subscription Table}: A cache-based hardware lookup table. 
\item \textbf{Subscription Buffer}: A fully-associative cache. 
\item Registers to accumulate the performance for each PIM core to dynamically turn subscriptions on/off.
\item Reserved space in memory to hold subscribed data.
\end{itemize}

For the \textit{central vault} in our \textit{global adaptive} policy (details will be discussed later in this section), we also added registers to record the performance for the entire system.

We modify the HMC packet transfer protocol, and add a new field to allow for the transfer of subscription request both alongside and separate from regular data access requests.

At early stages of our implementation, we have considered using a subscription threshold-based policy in which we have a count table to record the number of accesses to each address. The count table is a direct-mapped cache with 8192 (i.e., 8K) entries and 32-bits per entry, split between an 8-bit counter and a 24-bit tag. In each data access, we calculate a tag using the address of the block, and compare it with the tag of the corresponding entry in the count table. If the tag does not match, we reset the counter to 0 and update the tag with the current tag, by which, we evict current entry and replace it with the incoming entry. If the counter is greater than the threshold and the access is not local, the architecture will trigger the subscription routine. However, during our experiments, we discovered that almost all subscription-friendly workloads benefit from a 0-count threshold subscription policy in which we subscribe to an address on first access. Therefore, the count table is unnecessary and is not included in DL-PIM.

\noindent\textbf{Subscription Table.} We use a cache-based lookup table called the \textit{Subscription Table (ST)} which maps the original address of a data block to the current address with each subscription. The subscription table is a 4-way set-associative cache with 2048 sets per table per vault. Therefore, each vault has 8192 entries. Each entry has two addresses: the original address and one subscribed address. Additionally, each entry also has three state bits that indicate any of the following states:

\begin{itemize}
    \item Invalid (or "Unsubscribed")
    \item Pending Subscritpion
    \item Subscribed
    \item Pending Resubscription
    \item Pending Unsubscription
\end{itemize}

We discuss the use and meaning of each of these states in the next few subsections. 

\noindent\textbf{Subscription Buffer.} As our subscription table has limited size, we only allow for a subscription if space in the subscription table is available. When the subscription table is full, we utilize a replacement policy to locate the least frequently used or (in case of a tie) least recently used ST entry as our victim for \textit{unsubscription}. However, the unsubscription does not take effect instantly, and would require communication over the network. As such, the subscription buffer is used to temporarily store our request pending the completion of unsubscription.

The subscription buffer is a 32-entry fully-associative cache with an entry for subscription request that includes from vault, to vault, subscription address, and other information. The exact format of a subscription request will be provided in the following section. Each subscription buffer entry has a valid bit which will be set when the corresponding subscription table set has available space. In each cycle, 
%we check the valid bit of all entries, and 
we attempt to process a valid subscription request if any, based on valid bits.

\subsection{Subscription Protocol}

We implement a packet-based subscription protocol. Each subscription packet includes the following information:

\begin{itemize}
    \item From vault: The vault that originated the request.
    \item To vault: The destination vault for the request.
    \item Address: The memory address of the requested block.
    \item Request type: Possible request types are Subscription Request, Subscription Request Negative Acknowledgement, Subscription Data Transfer, Subscription Transfer Acknowledgement, Unsubscription Request, Unsubscription Transfer Acknowledgement, Turn On Subscription, Turn Off Subscription  
    \item Dirty bit: Whether the data being unsubscribed or resubscribed has been modified since subscription.
\end{itemize}

\subsubsection{Subscription Flow}
When an address is accessed, the vault that the subscription block will be subscribed to, called the "requester vault," first checks if it has the space to subscribe to the block. If space is unavailable, it pushes the subscription request into the subscription buffer to wait for unsubscription to free up space. Then, it pushes a packet with the address and sends it to the home vault that originally held the block, called the "original vault," and updates the subscription table of the corresponding address to the \textit{Pending Subscription} state.

Upon receiving the request, the original vault performs the same check, ensuring the address is not already subscribed or pending subscription/unsubscription to/from another vault, and updates the state to \textit{Pending Subscription}. Then, it starts the transfer of the subscribed data. The requester vault acknowledges the transfer. When both sides acknowledge the transfer, the subscription table entry is marked as \textit{Subscribed}.

\subsubsection{Resubscription Flow}
Our protocol includes a special subscription case called Resubscription, which occurs when a vault requests a subscription for a data block that is already subscribed to another location. In this scenario, the original vault redirects the request to the vault currently holding the memory block, known as the "subscribed vault." The subscribed vault then updates its subscription table entry to \textit{Pending Resubscription} and starts sending the data to the requester vault. Once the requester vault receives the data, it sends two acknowledgments: one to the original vault, allowing it to update its subscription table and change the entry to the new subscribed address, and one to the subscribed vault, allowing it to evict the entry from its table.

\subsubsection{Negative Acknowledgement of Subscription}
In some cases, a subscription request cannot be completed successfully. The subscription table has a limited size, and we use the subscription buffer to temporarily hold the address while making space. However, the subscription buffer is also limited in size. If the subscription buffer is full, we cannot complete the subscription. In this situation, the original vault sends a subscription negative acknowledgment to the requester vault, which then rolls back the subscription by removing the subscription entry. A similar situation occurs when the original vault (or the subscribed vault, in the case of resubscription) is currently in the process of subscribing the address block to another vault or unsubscribing the address block from another vault.

\subsubsection{Unsubscription Flow}
Unsubscription occurs when a vault's subscription table is full and needs to free up space. A special case arises when the requester vault is also the original vault, making it impossible to redirect the subscription request back to the original vault. In this case, the subscription request is converted into an unsubscription request.

Unsubscription can be initiated by either the original vault (wanting the data back) or the subscribed vault (wanting to return the data to the original vault). When the original vault initiates, it sends an unsubscription request to the subscribed vault. If the subscribed vault initiates, it skips this step. The subscribed vault then marks the relevant entry in its subscription table as \textit{Pending Unsubscription} and begins transferring the data back to the original vault. Once the original vault receives the data, it marks it as \textit{Unsubscribed} and acknowledges this to the subscribed vault, which also marks the entry as \textit{Unsubscribed}.

If an unsubscription request is made for an address that is in the process of being subscribed, the unsubscription process waits until the subscription flow is complete. Similarly, if an unsubscription request is made for an address undergoing resubscription, the process waits for the resubscription to finish and then forwards the unsubscription request to the requester vault to maintain data consistency.

\subsubsection{Dirty Bit}
To reduce the data transfer required for unsubscription, we add a dirty bit to our subscription reserved space, which is set when the block is written. During unsubscription, we check the status of the dirty bit. If the dirty bit is not set, we transfer only an acknowledgment packet instead of the full data, as the original vault already has the data. This dirty bit is also forwarded during resubscription using miscellaneous bits in the request packet.

\subsection{Memory Requests}

Unlike the baseline memory, our architecture employs a subscription table for dynamic address translation to any subscribed address. Because the subscription table is distributed across all vaults, any memory request necessitates that the vault requesting access ("requester vault") communicate with both the vault that originally held the block ("original vault") and the vault currently holding the most recent copy ("subscribed vault"). We will discuss the process of serving both read and write requests in the baseline memory implementation and our architecture.

We denote the Manhattan distance between the original vault and the requester vault as $h_{ro}$, between the requester vault and the subscribed vault as $h_{rs}$, and between the subscribed vault and the original vault as $h_{so}$. The size of a data block is $k-1$ flits, making a data transfer packet $k$ flits (with one flit used as a header for routing and other information).

Our memory access protocol is inspired by the SGI Origin system \cite{thesgiorigin} and based on the DASH protocol \cite{dashprotocol}. This protocol applies to the reserved areas in each vault containing subscribed blocks, in contrast to the caches used in the Origin system.

\noindent\textbf{Read Requests.} When serving a read request in the baseline memory implementation, the requester vault sends a request to the original vault. Upon receiving the request, the original vault transfers the data back to the requester vault. This process incurs a network overhead of $(k+1)h_{ro}$ cycles, assuming a single cycle latency per hop.

In our architecture, we first check if the requested address is in the local reserved subscription space. If it is, the access is local and incurs no network overhead. If the address is not in the local subscription space, we send the read request to the original vault, as the requester vault does not have up-to-date subscription information for the requested block. The original vault then checks its subscription table to identify the subscribed vault, if any. If the block is not subscribed to another vault, the original vault sends the data to the requester. If the block is subscribed, the original vault forwards the request to the subscribed vault, which then transfers the data back to the requester vault. Overall, a read access requires $h_{ro}+h_{so}+kh_{rs}$ cycles of network overhead.

\noindent\textbf{Write Requests.} In the baseline memory, the requester vault writes directly to the original vault, resulting in a network overhead of $kh_{ro}$ cycles.
In DL-PIM, for a remote access, the requester vault writes the data to the original vault. The original vault then forwards the written data to the subscribed vault. For a local access, there is no network overhead.

\subsection{Adaptive Subscription Architecture}
Our implementation supports both binary \textit{always-subscribe} and \textit{never-subscribe} configurations. While many workloads benefit from proactively subscribing and moving data to the requester vault, others suffer from the additional latency and traffic caused by the indirection requests described in the previous section. Therefore, we also implemented and evaluated several adaptive strategies to dynamically enable and disable subscriptions. Our adaptive architectures focus on two key properties: the number of hops traveled and the access latency of each request. Based on these properties, we designed the Hops-based Adaptive Policy and the Latency-based Adaptive Policy, which we will discuss in the following sections.

\subsubsection{Adaptive Policy Implementation}
We aggregate the necessary statistics for our policies in hardware registers, as illustrated in Figure~\ref{fig:hardware_structures}. 

Initially, we considered a policy that continuously tracked the cost and benefit of subscription in a \textit{feedback register} (discussed below), and broadcast policy changes to every vault whenever the feedback register's value shifted from positive to negative or vice versa. However, this proved inefficient because the subscription policy changed rapidly, and broadcasting the new policy to all vaults incurred high overhead.

To address this, we divided the execution into "epochs" and made decisions based on the collected information at the end of each epoch, which in our implementation is every $10^6$ (one million) cycles. At the end of each epoch, a decision about the subscription policy for the next epoch is broadcast, and the information stored in the aggregate registers is cleared, ensuring that each epoch's decision is based solely on the previous epoch's cost/benefit behavior.

\subsubsection{Hops-based Adaptive Policy}
Since our architecture attempts to reduce data movement across the memory network, the number of hops per packet is a key metric to consider. To assess whether subscriptions are helping or hurting traffic, we use a register called \textit{feedback register} at each vault. We measure the number of hops travelled by each packet, and using the vault's address to estimate the number of hops travelled if the requested address is not subscribed. We then compare the two hops' counts. If the estimated original hop is higher than the actual number of hops when subscribed, it means that our subscription helps reduce the distance traveled for that packet, and we therefore increment the feedback register. Otherwise, if the subscribed hops is higher, it means that our subscription is increasing the distance traveled per packet, and we therefore decrement the feedback register. As such, a positive feedback register value means that the workload is benefiting from subscription, while a negative value means that the workload is negatively impacted. In the first epoch, we turn on subscription across all vaults and collect the cost/benefit information in each vault's feedback register. At the end of each epoch, if the feedback register is negative, we turn off the subscription for the next epoch; otherwise we turn it on. 

\subsubsection{Latency-based Adaptive Policy}
A more accurate predictor of performance, though harder to collect at runtime, is the access latency per request. This metric includes all latency components, such as transfer (communication) and queuing delays. While estimating hops without subscription is straightforward, as discussed above, latency depends on factors beyond the number of hops traveled, making it difficult to estimate for a memory request without subscription.

To address this, we used a different strategy for making subscription decisions. At the completion of each request, we record its latency by adding it to a \textit{latency register} (as shown in Figure~\ref{fig:hardware_structures}). We also record the number of requests served in a given epoch in another register called the \textit{request register}. At the end of each epoch, we calculate the average latency for all memory requests during that epoch.

In the initial epochs, we use the hops-based feedback register to decide the subscription policy and aggregate the epoch's request latency into the latency register. For each subsequent epoch, we compare the average latency per memory request to that of the previous epoch, stored in a \textit{previous latency register}. If the average latency is lower or within a certain threshold, we maintain the current subscription policy for the next epoch. If the per-request average latency increases beyond the threshold, we reverse the decision and enable or disable subscription in the next epoch.

After experimenting with various thresholds, we found that a 2\% threshold is heuristically optimal for most workloads. Thus, we use a 2\% latency change threshold for the latency-based adaptive policy.

\subsubsection{The \textit{Subscription Away} Problem and Solutions}
Both hops-based and latency-based adaptive policies suffer from the "subscription away" problem. While the feedback register mechanism assesses the benefits of subscription for a vault, it overlooks the negative impact on the original vault from which the data was subscribed. To address this, we update both the accessing and subscribed vaults' feedback registers when negative feedback occurs due to increased hops for memory requests.

Similarly, the latency-based policy faces the same issue, as some vaults benefit at the expense of others. From the perspective of the subscribed vault, the average latency appears lower since the subscribed block incurs only local access latency, while other vaults bear communication and queuing delays. However, the negative impact on other vaults is not accounted for by the feedback mechanism of the subscribed vault. Without negative feedback, all vaults might opt for the greedy policy of \textit{always-subscribe}, potentially leading to thrashing and overall performance and energy degradation.

To address this problem, decisions are made globally. Each vault sends a special packet to the central vault positioned at the center of the network before an epoch ends. The central vault then calculates global latency averages, decides on the subscription policy, and broadcasts this decision to every vault. However, this complex computation introduces some latency (estimated to be around 1000 cycles) before the new global policy takes effect across all vaults at the start of an epoch.

\subsubsection{The \textit{Always Unsubscription} Problem and Solutions}
Another issue with the initial adaptive policy arises because we can only compute the number of hops for subscribed requests when subscription is enabled. Without subscriptions, it's challenging to determine when subscription would enhance performance, potentially leaving the adaptive policy stuck at the \textit{never-subscribe} setting. While the latency-based adaptive policy may indicate incorrect shifts to \textit{never-subscribe} due to higher latency, transitioning between phases where subscription benefits fluctuate remains a challenge.

To address this, we implemented a dynamic set sampling mechanism similar to one proposed by Qureshi et al.~\cite{mlpawarecacherep}. This mechanism involves designating two \textit{leading sets}: one with subscriptions always enabled and the other with them always disabled. Feedback and latency are recorded separately for each leading set. At the end of each epoch, we compare hops and/or average latency for the leading sets to determine the most beneficial subscription strategy. If one leading set shows lower latency or higher feedback, we adopt its policy for all sets in our subscription table in the next epoch.

While this resolves the "always unsubscription" problem, it may cause some memory locations to consistently remain in the \textit{never-subscribe} set used in set sampling. This can result in corner cases where data blocks that would benefit from even access distribution will be assigned to \textit{never-subscribe} , resulting in degraded performance for workloads that benefit from evenly distributed demand across vaults, such as CHABsBez and SPLRad. As the placement of memory location is hard to predict at runtime, this would diminish 
%Runtime memory location prediction challenges may lead heavily accessed locations to fall within the \textit{never-subscribe} set, diminishing 
potential benefits from our adaptive DL-PIM architecture.

\section{Evaluation Methodology}
\label{sec_evaluation}

\subsection{Experimental Setup}
We employ the DAMOV simulation framework~\cite{oliveira2021damov} to implement and evaluate our proposed architecture. DAMOV seamlessly integrates the ZSim~\cite{zsim} CPU simulator and the Ramulator~\cite{ramulator} memory simulator. This framework allows us to simulate a configurable number of traditional CPU cores or Processing-in-Memory (PIM) cores with various memory technologies, including HMC and HBM.

\begin{figure}[t]
\centering
\includegraphics[width=0.45\textwidth]{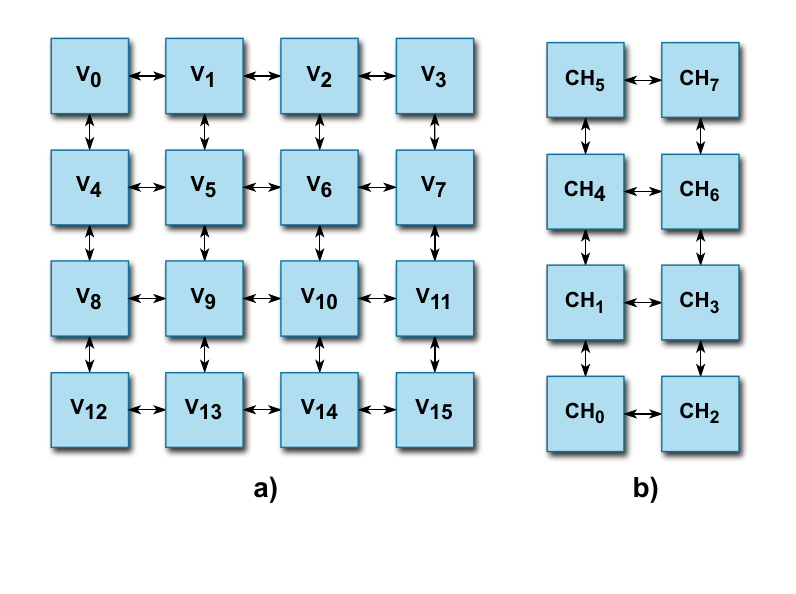}
\vspace{-25pt}
\caption{Internal network model a) Representation of 32 vaults in a $6\times6$ HMC network. b) Representation of 8 channels in a $4\times2$ HBM network.}
\vspace{-5pt}
\label{fig:network}
\end{figure}

\textbf{Network model}. We simulate an inter-vault network model for a $6\times6$ interconnected memory network with 32 vaults, as shown in Figure~\ref{fig:network}.a for HMC memory, and a $4\times2$ network for HBM memory, as depicted in Figure~\ref{fig:network}.b. We have instrumented DAMOV to dynamically analyze inter-vault traffic and overhead in a distributed PIM system.

\textbf{Baseline configuration}. In our evaluation, we configure a baseline system similar to \cite{oliveira2021damov} that has 2.4Ghz PIM cores with 32KB L1 cache directly connected to a 4GB\_HMC/4Gb\_HBM memory system. Table~\ref{table:config_hmc} and Table~\ref{table:config_hbm} includes detailed configurations of the simulated memory baseline systems.

\begin{table}[t]
\caption{Simulated HMC Baseline System configuration.}
\label{table:config_hmc}
\vspace{-5pt}
\small
\begin{tabular}{|c|c|}
\hline
OS & Ubuntu 18.04 x86-64 running on QEMU 4.2.1\\ \hline
CPU& two Intel(R) Xeon(R) Gold 6248R CPU @ 3.00GHz\\ \hline
\multirow{4}{*}{Memory} & 128GB total size; HMC v2.0; 32 vaults;\\
&8 DRAM banks/vault; 256B row buffer; \\
% &DRAM@166 MHz; \\
&8B burst width at 2:1 core-to-bus freq. ratio; \\
&Open-page policy; HMC default interleaving~\cite{oliveira2021damov} \\
\hline

\end{tabular}
\vspace{-3pt}
\end{table}

\begin{table}[t]
\caption{Simulated HBM Baseline System configuration.}
\label{table:config_hbm}
\vspace{-5pt}
\small 
\begin{tabular}{|c|c|}
\hline
OS & Ubuntu 20 x86-64\\ \hline
CPU& two Intel(R) Xeon(R) Gold 6248R CPU @ 3.00GHz\\ \hline
\multirow{2}{*}{Memory} & 4GB total size; HBM2; 8 channels;\\
& 4 bank-groups/channel; 4 banks/bank-group \\
\hline
\end{tabular}
\vspace{-5pt}
\end{table}

\textbf{Workloads}. The DAMOV benchmark suite included over 300 applications from different domains such as big data and machine learning ~\cite{oliveira2021damov}. We use the 31 \textit{representative} applications identified by DAMOV for our evaluation. We did not select short applications to provide useful insights. Those representative applications are from benchmark suites Chai~\cite{chai}, Darknet~\cite{darknet13}, Hashjoin~\cite{hashjoin}, Ligra~\cite{ligra}, phoenix~\cite{phoenix}, PolyBench~\cite{polybench}, Rodinia~\cite{rodinia}, SPLASH2~\cite{splash2} and STREAM~\cite{mccalpin1995memory}. Table~\ref{table:benchmarks} displays functions alongside their corresponding benchmarks utilized in our simulation from these benchmark suites. The final column indicates our assigned abbreviated names for ease of reference.

 % In the next section, we present results from these applications and focus on 13 that have non-negligible data reuse for most of our analysis (Table~\ref{table:benchmarks}). A full list of all workloads is shown in Table~\ref{table:Execution_time} and their description is available in~\cite{oliveira2021damov}.

\textbf{Simulation Experimental Methodology}. For our simulation results, we use $10^6$ memory requests to warm up our caches and hardware structures, and run the remainder of the workload to generate results. Additionally, to mitigate runtime variability, each workload/configuration is simulated five times, and the average of these runs is presented in our figures.

\begin{table}[t]

\centering
\fontsize{8.5}{10.2}\selectfont
\setlength{\tabcolsep}{2pt}
\caption{Simulated workloads.}
\label{table:benchmarks}
\vspace{-5pt}
\renewcommand{\arraystretch}{1.2}
\centering
\begin{tabular}{|c|p{2.5cm}|p{2.8cm}|p{1.6cm}|}
\hline
Suite & Benchmark & Function & Short Name \\ 
\hline
\multirow{2}{*}{Chai} & Bezier Surface & Bezier & CHABsBez \\ \cline{2-4}
 & Padding & Padding & CHAOpad \\ \hline
Darknet & Yolo & gemm\_nn & DRKYolo \\ \hline
\multirow{2}{*}{Hashjoin} & NPO & ProbeHashTable & HSJNPO \\ \cline{2-4}
 & PRH & HistogramJoin & HSJPRH \\ \hline
\multirow{5}{*}{Ligra} & Betweenness Centrality & \multirow{3}{*}{EdgeMapSparse (USA)} & LIGBcEms \\ \cline{2-2} \cline{4-4}
 & Breadth-First Search & & LIGBfsEms \\ \cline{2-2} \cline{4-4}
 & BFS-Connected Components & & LIGBfscEms \\ \cline{2-4}
 & PageRank & EdgeMapDense (USA) & LIGPrkEmd \\ \cline{2-4}
 & Triangle & EdgeMapDense (Rmat) & LIGTriEmd \\ \hline
Phoenix & Linear Regression & linear regression map & PHELinReg \\ \hline
\multirow{8}{*}{PolyBench} & \multirow{6}{*}{Linear Algebra} & 3 Matrix Multiplications & PLY3mm \\ \cline{3-4}
 & & Multi-resolution analysis kernel & PLYDoitgen \\ \cline{3-4}
 & & Matrix-multiply C=alpha.A.B+beta.C & PLYgemm \\ \cline{3-4}
 & & Vector Mult. and Matrix Addition & PLYgemver \\ \cline{3-4}
 & & Gram-Schmidt decomposition & PLYGramSch \\ \cline{3-4}
 & & Symmetric matrix-multiply & PLYSymm \\ \cline{2-4}
 & \multirow{2}{*}{Stencil} & 2D Convolution & PLYcon2d \\ \cline{3-4}
 & & 2-D Finite Different Time Domain & PLYdtd \\ \hline
\multirow{2}{*}{Rodinia} & BFS & BFSGraph & RODBfs \\ \cline{2-4}
 & Needleman-Wunsch & runTest & RODNw \\ \hline
\multirow{6}{*}{SPLASH2} & \multirow{2}{*}{FFT} & Reverse & SPLFftRev \\ \cline{3-4}
 & & Transpose & SPLFftTra \\ \cline{2-4}
 & \multirow{2}{*}{Oceanncp} & jacobcalc & SPLOcnpJac \\ \cline{3-4}
 & & laplaccalc & SPLOcnpLap \\ \cline{2-4}
 & Oceancp & slave2 & SPLOcpSlave \\ \cline{2-4}
 & Radix & slave\_sort & SPLRad \\ \hline
\multirow{4}{*}{STREAM} & Add & Add & STRAdd \\ \cline{2-4}
 & Copy & Copy & STRCpy \\ \cline{2-4}
 & Scale & Scale & STRSca \\ \cline{2-4}
 & Triad & Triad & STRTriad \\ \hline
\end{tabular}
%\vspace{-10pt}
\end{table}

% \subsection{Baseline analysis}
% We simulated the baseline configuration without any subscription and the \textit{always-subscribe} configuration where we always subscribe on first access to a memory block. Figure~\ref{fig:baseline_mpki} shows the average absolute misses per thousand instructions for all workloads. We show normalized performance gain for the remainder of this section. 
% As shown in this Figure, each workload has different memory access patterns and execution characteristics, and therefore performs differently with our architecture.

% \begin{figure}[ht]
% \centering
% \includegraphics[width=0.45\textwidth]{images/baseline_mpki.png}
% \caption{Baseline MPKI without warmup.}
% \label{fig:baseline_mpki}
% \end{figure}

\subsection{Simulation Results}
\label{sec_results}
\subsubsection{\textit{Always-Subscribe} Policy Analysis}
\begin{figure}[t]
% \vspace{-5pt}
\centering
\includegraphics[width=0.48\textwidth]{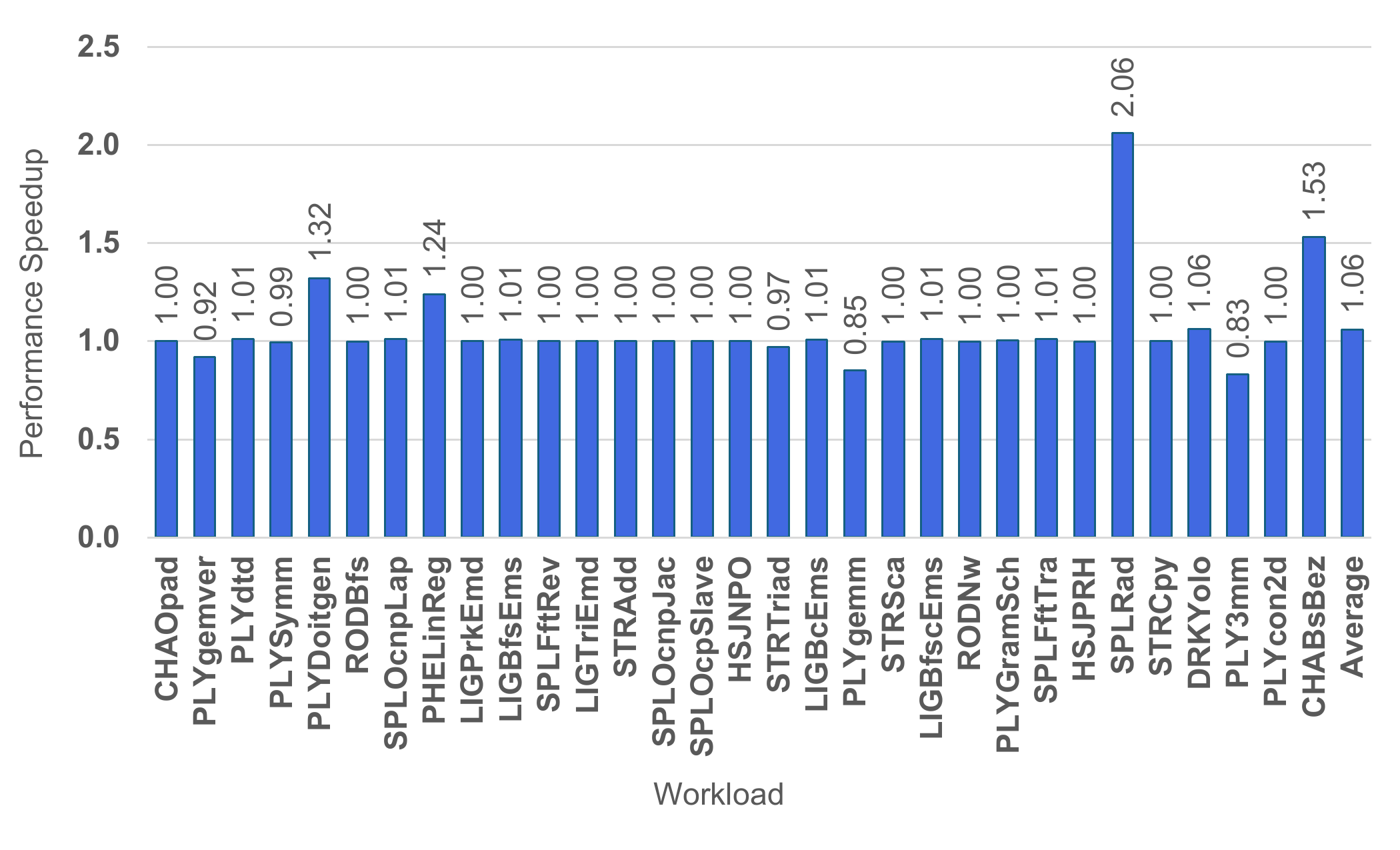}
\vspace{-5pt}
\caption{Performance gain for the \textit{always-subscribe} policy, measured by the execution cycles of the baseline divided by that of the \textit{always-subscribe} policy.}
%\vspace{-8pt}
\label{fig:normalized_cycles_allocate_1h0c}
\end{figure}

In Figure~\ref{fig:normalized_cycles_allocate_1h0c}, we analyzed the performance of all 32 representative workloads with the \textit{always-subscribe} policy, where we always perform subscription on the first access of a memory location. Some workloads show significant speedups. For example, SPLRad have up to 105\% performance gain. Conversely, workloads such as PLYgemm and PLY3mm have up to 17\% performance loss. On average, all benchmarks have a performance gain of nearly 6\%.

\begin{figure}[t]
% \vspace{-10pt}
\centering
\includegraphics[width=0.48\textwidth]{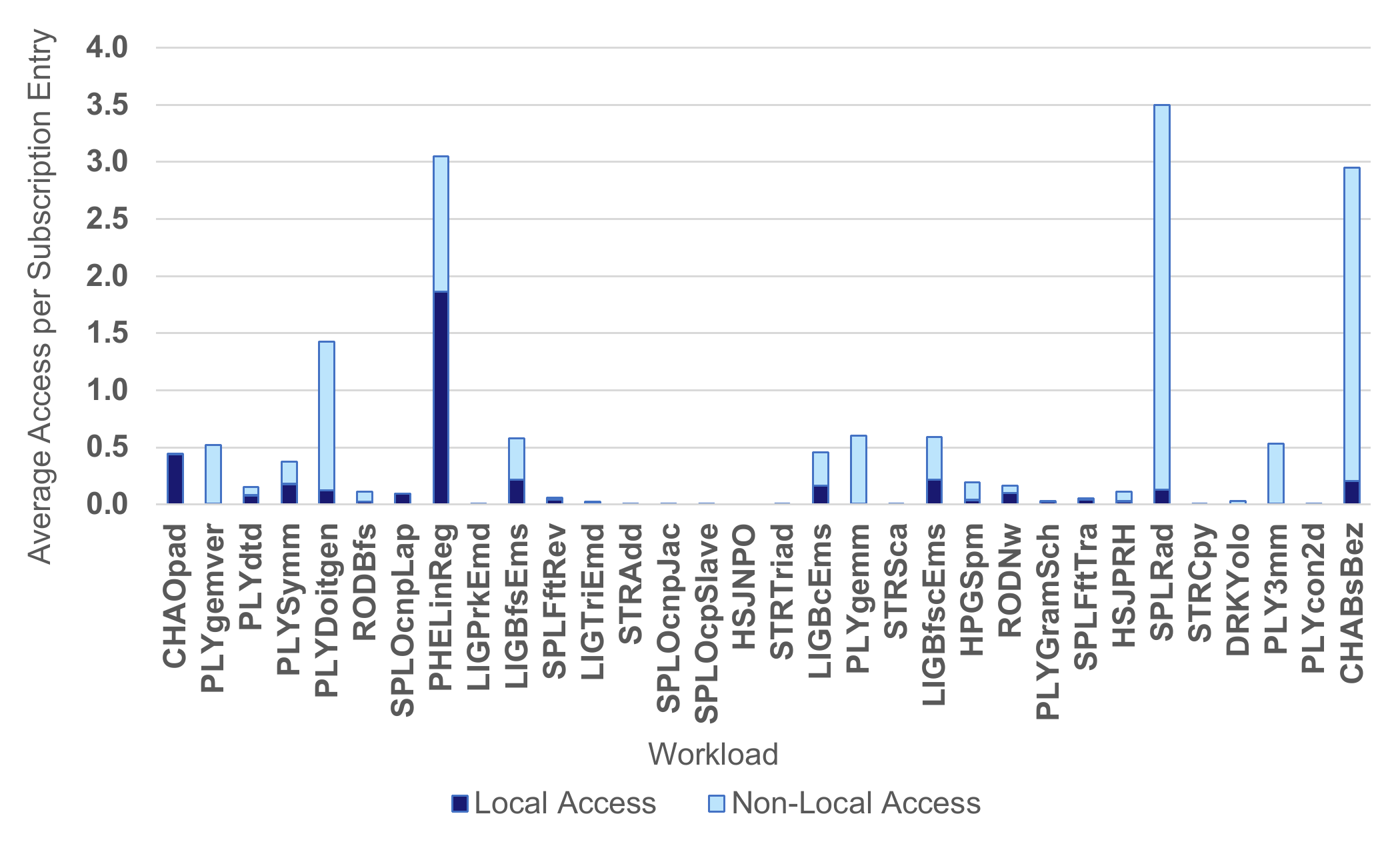}
\vspace{-10pt}
\caption{Average number of local and non-local accesses per subscription of \textit{always-subscribe}. The dark blue portion indicates the average number of local accesses from the subscribed vault to a subscribed block after it was successfully subscribed. The light blue portion indicates the average number of remote accesses to a subscribed block. }
\vspace{-5pt}
\label{fig:avg_sub_access_breakdown_allocate_1h0c}
\end{figure}

This Figure also shows that many workloads do not demonstrate any performance impact (Speedup 1.00), indicating subscription has little to no effect. This is despite the fact that for many of these benchmarks, data transfer and queuing delays still represent a considerable percentage of overall memory latency (Figure~\ref{fig:memory_latency_breakdown}). One reason for this minimal impact is the poor reuse properties of many workloads as shown in Figure~\ref{fig:avg_sub_access_breakdown_allocate_1h0c}. This figure shows the average number of times a subscribed block is reused, either by the local subscribed vault (dark blue) or by a remote vault (light blue). Many workloads display near-zero average reuse per block, meaning blocks are seldom accessed again after being subscribed and moved to the subscribed vault. This does not incur additional overhead for the \textit{always-subscribe} policy since the subscription data transfer would have occurred in the baseline when a processing unit requests a block from a remote vault. However, the lack of reuse means that no performance benefit is realized from \textit{always-subscribe}. Therefore, \textbf{in the remainder of the paper, we focus only on workloads that show non-negligible reuse.}

\begin{figure}[t]
\vspace{-20pt}
\centering
\includegraphics[width=0.48\textwidth]{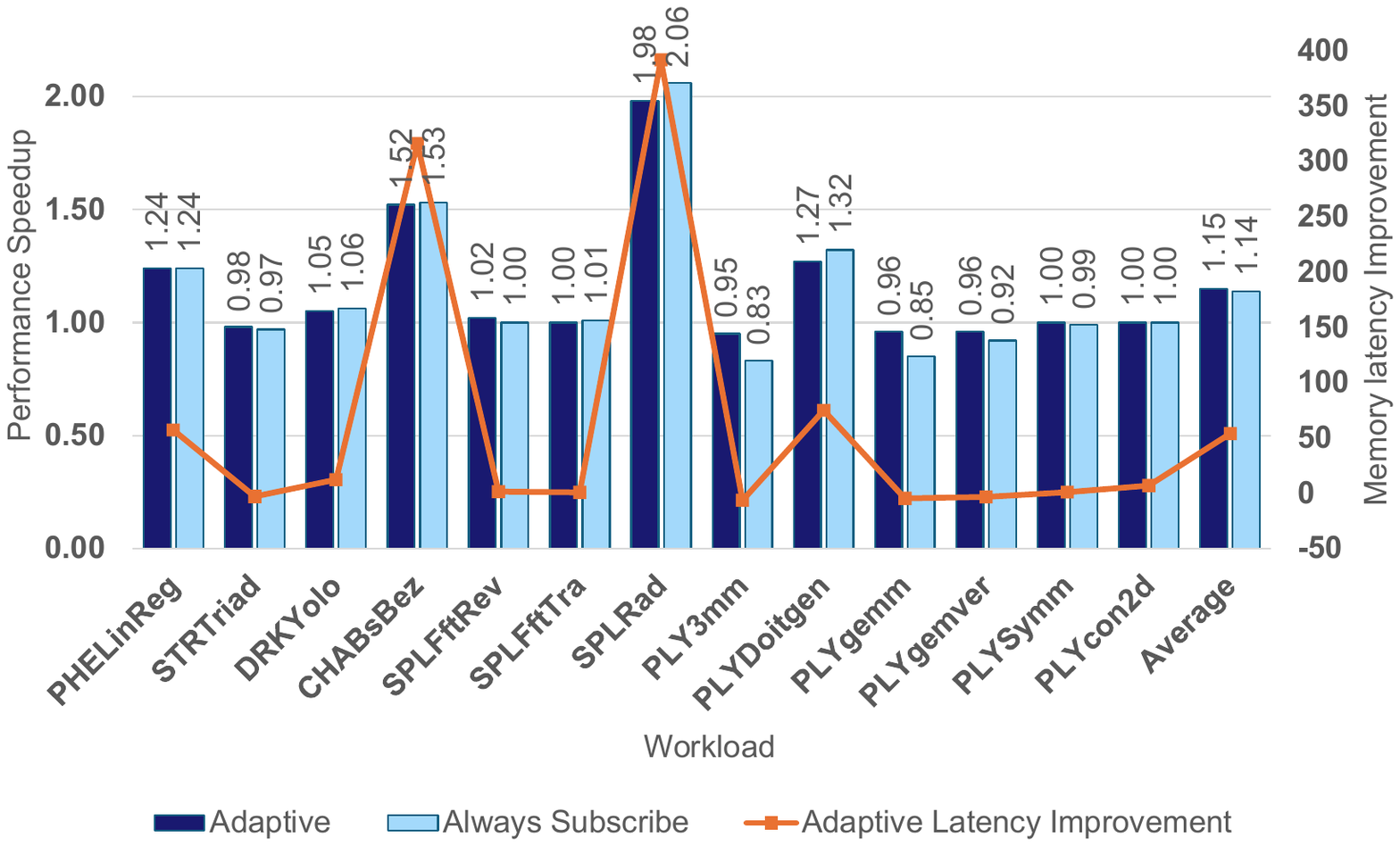}
\vspace{-20pt}
\caption{Performance gain of \textit{always-subscribe} and \textit{adaptive} normalized to the baseline (Bars). Memory latency improvement (Orange line).}
\label{fig:perf-adaptive}
\vspace{-10pt}
\end{figure}

\subsubsection{\textit{Adaptive} Policies Analysis}

Figure~\ref{fig:perf-adaptive} shows that both \textit{always subscribe} and \textit{adaptive} policies affect performance significantly for many workloads. While most selected workloads benefit from \textit{always-subscribe}, some are negatively affected. The average of speedup for selected benchmarks is nearly 14\% for \textit{always-subscribe} and 15\% for \textit{adaptive}, which successfully reduces performance degradation for workloads that are hurt by \textit{always-subscribe}.

To provide further insight into the benefits of the subscription method, we show the average memory latency improvement for the adaptive method compared to the baseline, represented by the orange line in Figure~\ref{fig:perf-adaptive}. As the figure illustrates, workloads with greater latency improvements achieve higher performance speedup. On average, the adaptive policy reduces the average memory latency per request by nearly 54\% across the selected workloads.

\begin{figure}[t]
\vspace{-15pt}
\centering
\includegraphics[width=0.48\textwidth]{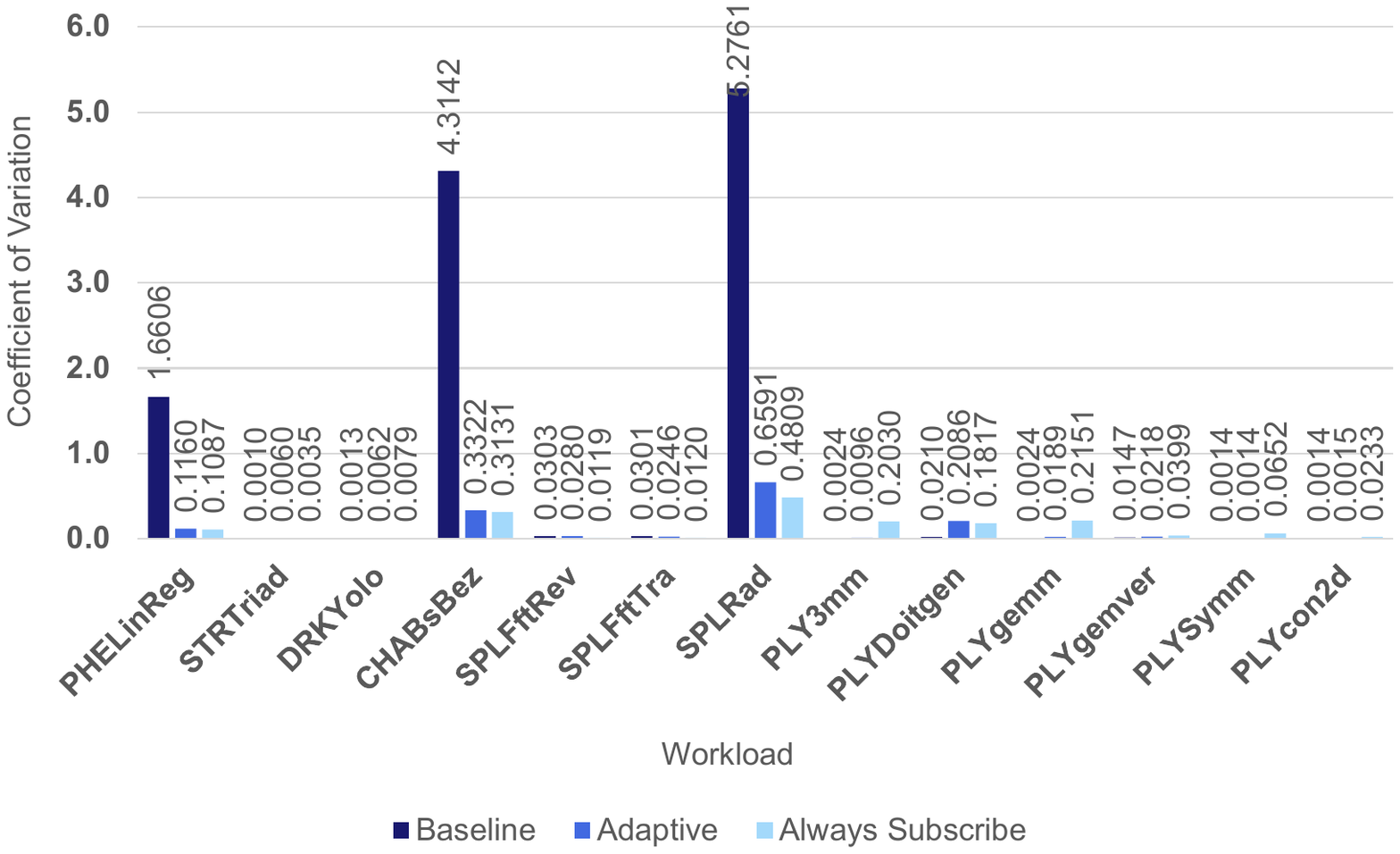}
\vspace{-31pt}
\caption{Coefficient of variation of the access distribution per vault for \textit{always-subscribe} and \textit{adaptive} policies. A high CoV implies uneven distribution where some vaults have much higher demand than others.}
\label{fig:cov-adaptive}
\vspace{-5pt}
\end{figure}

\begin{figure}[t]
%\vspace{-7pt}
\centering
\includegraphics[width=0.48\textwidth]{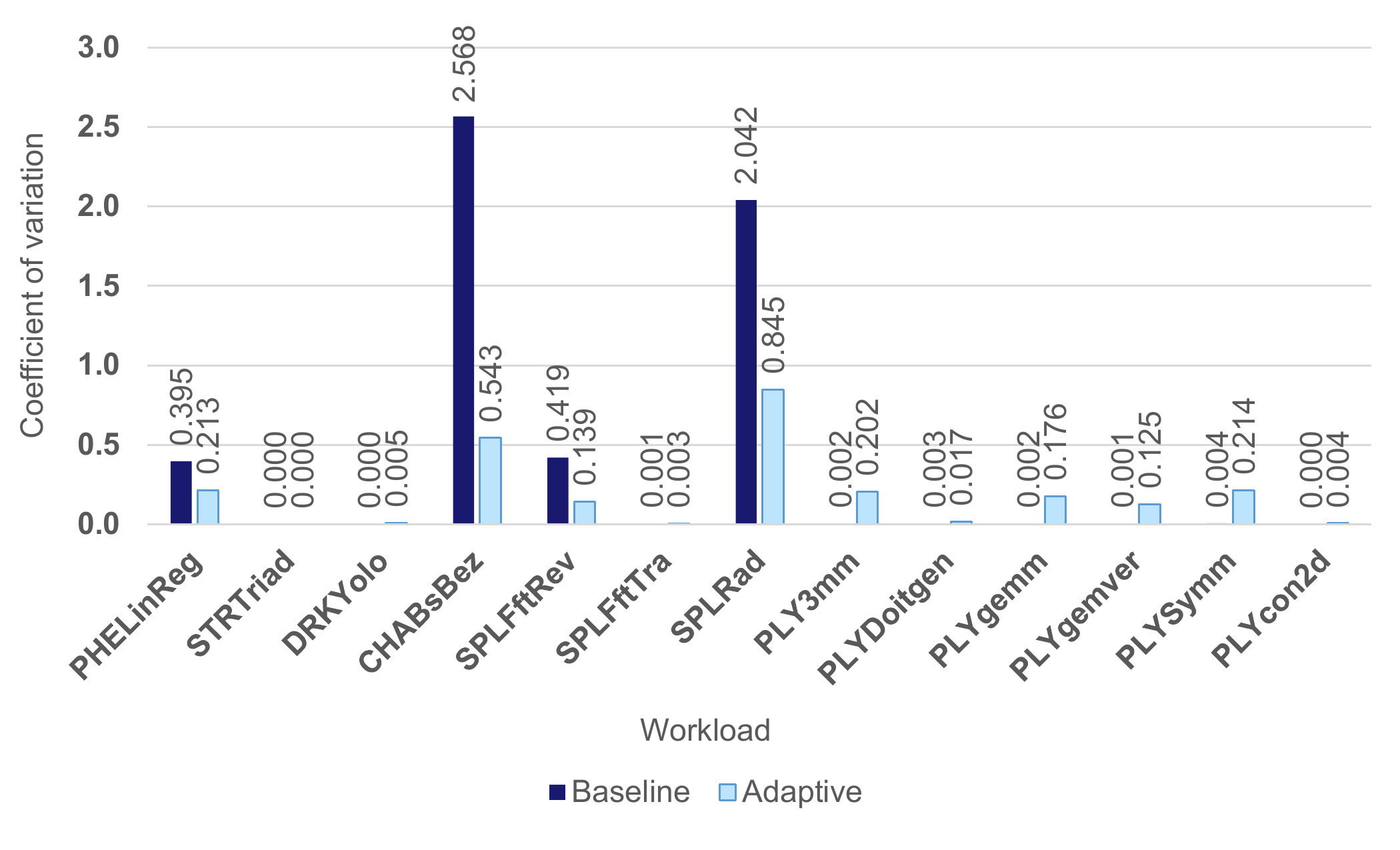}
%\vspace{-13pt}
\caption{CoV of memory access distribution for \textit{baseline} and \textit{adaptive} policies in HBM.}
\label{fig:cov-adaptive_hbm}
\vspace{-10pt}
\end{figure}

Additionally, our method helps reduce the CoV of access distribution per vault for workloads with high CoV, such as PHELinReg, CHABsBez, and SPLRad, which had the highest CoV among our workloads as shown in Figure~\ref{fig:cov-adaptive} and Figure~\ref{fig:cov-adaptive_hbm}. This indicates that our approach effectively balances the distribution of memory accesses across vaults.

\begin{figure}[t]
\centering
\vspace{3pt}
\includegraphics[width=0.48\textwidth]{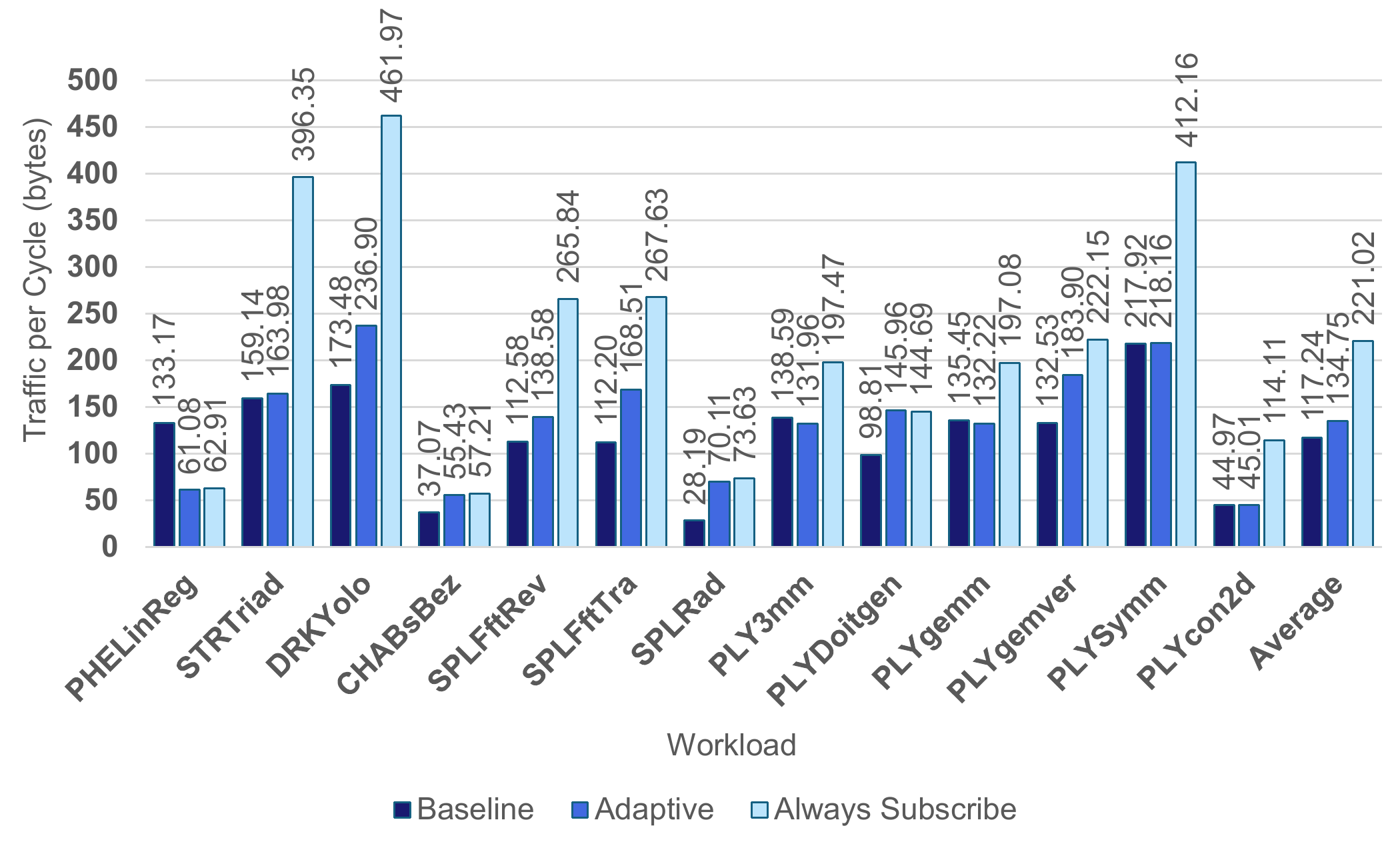}
\vspace{-5pt}
\caption{Average network traffic (in bytes per cycle) for \textit{always-subscribe} and \textit{adaptive} policies.}
\vspace{-10pt}
\label{fig:traffic-adaptive}
\end{figure}

Another interesting metric is total traffic, which summarizes the total network bandwidth demand. Figure~\ref{fig:traffic-adaptive} shows traffic per cycle for \textit{always-subscribe} and \textit{adaptive} compared to the baseline, including traffic from both memory accesses and subscription requests. Some workloads, like PHELinReg, significantly reduce bandwidth demand compared to the baseline. However, most workloads exhibit higher bandwidth demand with \textit{always-subscribe}, averaging an 88\% increase over the baseline due to additional subscription traffic, mainly caused by the low degree of reuse for subscribed blocks (Figure~\ref{fig:avg_sub_access_breakdown_allocate_1h0c}). In contrast, the \textit{adaptive} policy results in an average increase of only 14\%. Although both DL-PIM policies increase average network bandwidth, the reduction in execution time from more local accesses contributes to a more energy-efficient design.

Therefore, while the \textit{always-subscribe} policy offers substantial performance benefits for many workloads, it also introduces significant bandwidth overhead and can negatively impact certain workloads. The \textit{adaptive} policy, however, successfully balances performance gains with lower increases in bandwidth demand, offering a more efficient solution. By dynamically adjusting subscription policies based on workload behavior, the \textit{adaptive} policy reduces performance degradation and network bandwidth demand, leading to a more energy-efficient memory access strategy in DL-PIM architectures.

\begin{figure}[t]

\centering
\includegraphics[width=0.48\textwidth]{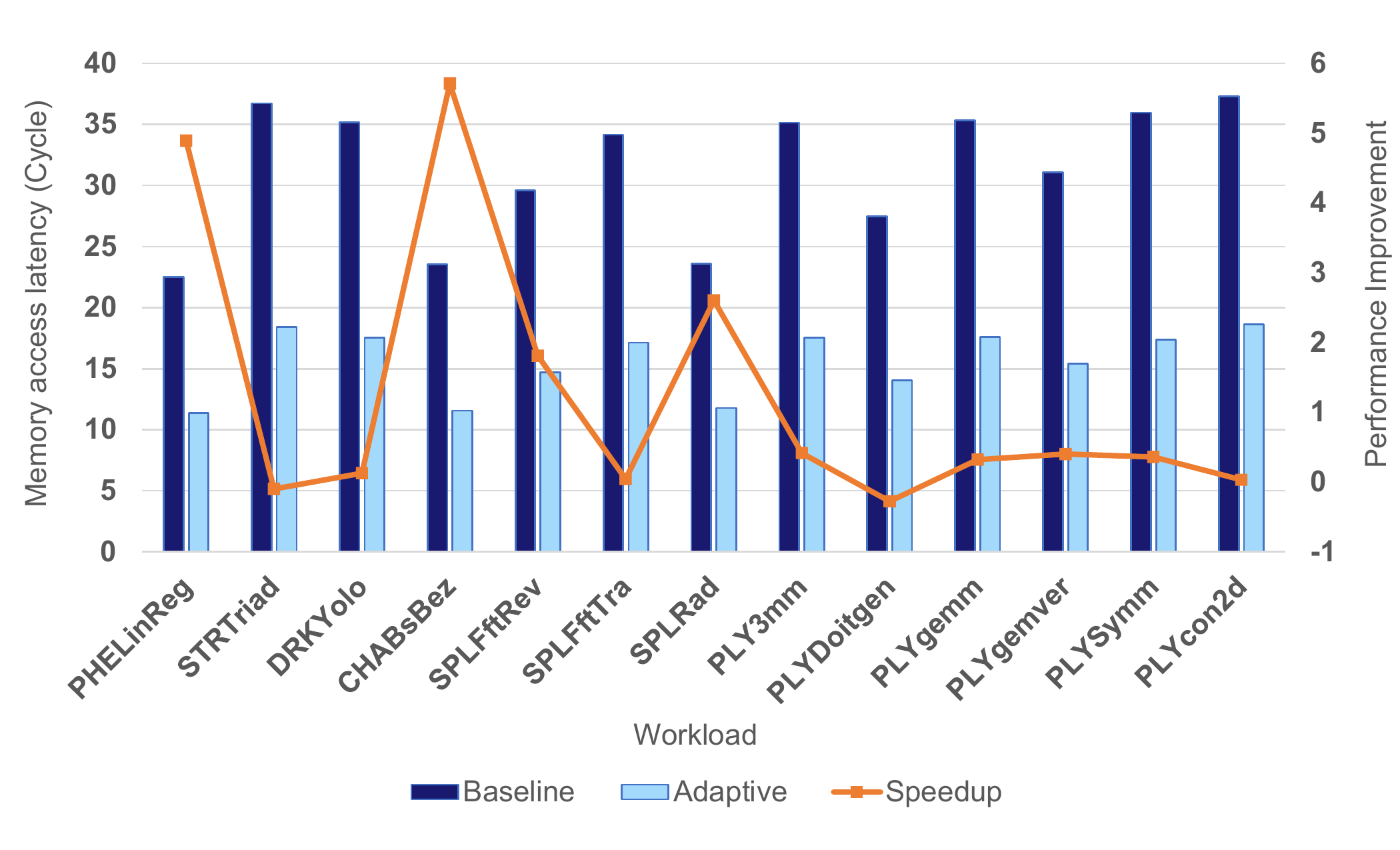}
%\vspace{-15pt}
\caption{Memory access latency comparison between baseline and adaptive policy in HBM (Bars). Speedup Improvement Percentage (Orange line).}
%\vspace{-5pt}
\label{fig:hbm_latency}
\end{figure}

We also evaluated the adaptive policy on HBM systems. Figure~\ref{fig:hbm_latency} presents a comparison of memory access latency between the baseline and adaptive policies. The results indicate that the adaptive subscription policy effectively decreased memory access latency across workloads, demonstrating its efficiency in optimizing memory performance in HBM systems. On average, latency was reduced by about 50\%. This improvement is due to the more even distribution of memory requests across the available channels, reducing contention and improving access times. Additionally, keeping subscribed data locally accessible reduces overall demand on memory bandwidth. The lower complexity of HBM's network model (relative to HMC's network) further enhances the effectiveness of the adaptive subscription policy. 

The performance improvements for different workloads are also noteworthy. Workloads such as CHABsBez, SPLRad, and PHELinReg show the most significant improvements, due to their high coefficients of variation as shown in Figure~\ref{fig:cov-baseline}. This indicates an uneven distribution where some
vaults have much higher demand than the rest, therefore
dominating performance. Similar improvements can also be seen in HMC, as illustrated in Figure~\ref{fig:perf-adaptive}.

\subsection{Sensitivity to Subscription Table Sizes}

\begin{figure}[t]
\centering
\includegraphics[width=0.48\textwidth]{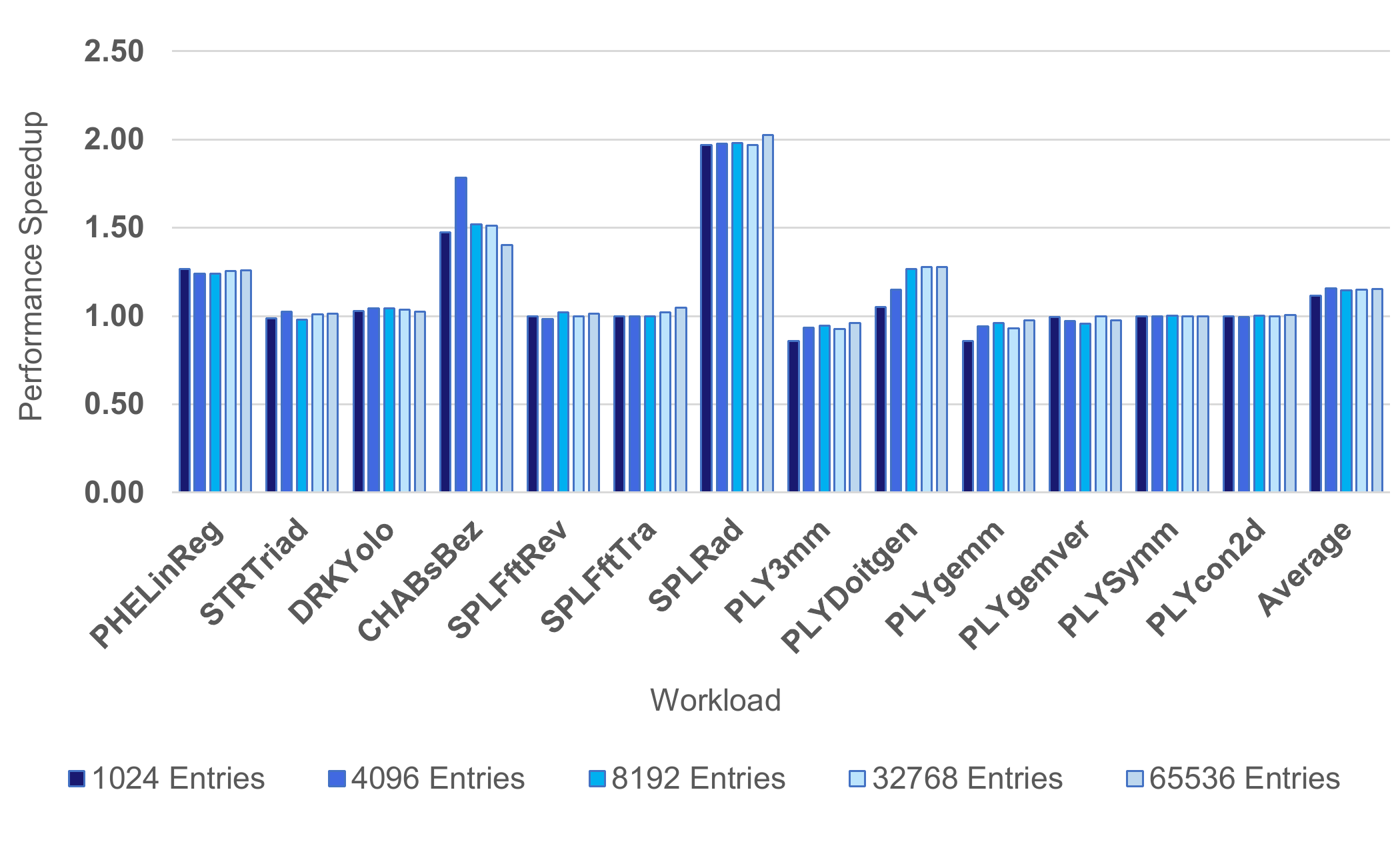}
%\vspace{-15pt}
\caption{\textit{Adaptive} speedup with different subscription table sizes.}
%\vspace{-6pt}
\label{fig:sensitivity-tablesizes}
\end{figure}

We analyzed the impact of subscription table sizes on the performance. With larger tables, DL-PIM can hold more subscribed memory locations for each vault, but would increase hardware overhead. Figure~\ref{fig:sensitivity-tablesizes} shows that some workloads like PLYDoitgen saw a performance improvement with a larger subscription table, but that improvement flattened with a 8192-entry subscription table. We decided to use 8192 entries as our default configuration in this paper, which incurs a 0.125\% state overhead relative to the 4GB vault memory size. 

%We analyzed the impact of subscription table sizes on performance. Larger tables allow DL-PIM to store more subscribed memory locations per vault but come with increased hardware overhead. As shown in Figure~\ref{fig:sensitivity-tablesizes}, workloads like PLYDoitgen exhibited performance improvements with larger subscription tables, though these improvements flattened with an 8192-entry table. Consequently, we chose 8192 entries as our default configuration in this paper, resulting in a state overhead of just 0.125\% relative to the 4GB vault memory size.  Why is this repeated?

\section{Related Work}
\label{sec_related_work}

\textbf{Stacked DRAM latency reduction:} Previous research~\cite{fataglessdramcache,mostlycleandramcache} has proposed using stacked DRAM as a memory-side cache, which automatically copies or moves data upon request. Most DRAM cache studies aim to increase capacity or reduce latency~\cite{fundamentallattradeoff,enableconventionalblocksizes,accord,effectivenessof3dstackedcaches,atcachesramtagcache,bimodaldramcache,banshee}, but these efforts primarily focus on processor-centric systems. Other studies aim to reduce communication latency between distributed DRAM caches in multi-node systems~\cite{candydramcaches}. In contrast, our proposal leverages stacked DRAM technology in a PIM system to minimize data movement between the processor and memory.

\textbf{Stacked DRAM network:} Previous work~\cite{memnet} has proposed replacing the crossbar network in HMC systems with a mesh network and a reduction/dispersion tree architecture, where memory controllers in the same row are interconnected via a set of reduction /dispersion trees. While this approach reduces communication overhead between memory controllers and external processors, it is not well-suited for PIM architectures. The primary limitation is that it restricts direct communication between memory controllers without routing through the processor, thereby failing to reduce processor-memory data movement.

\textbf{Network overhead reduction:} Tian et al.~\cite{tian2023abndp} (ABNDP) proposed a method for reducing network overhead in PIM systems through a hardware/software co-design. This approach introduces a new programming model and employs a \textit{traveller cache} for remote data, which functions similarly to a traditional cache. A single traveller cache is shared among four or more vaults. ABNDP breaks programs into smaller tasks using its programming model and dynamically schedules these tasks based on data location. In contrast, our DL-PIM proposal is a hardware-only solution utilizing a 3D-stacked memory system. It employs a reserved area in memory, rather than a separate cache, to subscribe to remote addresses and minimize data movement for each memory access.

\textbf{Subscription-based framework:} GPS~\cite{muthukrishnan2021gps} employs a subscription-based hardware-software integrated framework to accelerate multi-GPU systems. It utilizes manual subscription management in software, enabling data usage by both local and remote GPUs, and evaluates several automatic subscription algorithms. While inspired by this work, DL-PIM is a hardware-only solution that preserves compatibility with existing program models and APIs.

\textbf{Cache-only Memory Access: } Another architecture that  has some similarities with DL-PIM at a high-level is Cache-only Memory Access (COMA) architecture~\cite{coma,comparativeofccnumaandcoma,reactivenuma}. COMA dynamically migrates data requested by a local node to a local \textit{attraction memory} from a remote node upon access. However, COMA can create multiple copies of the same data block across different attraction memories for shared blocks. In contrast, our proposal invalidates the original copy upon subscription. While this might lead to performance degradation, it simplifies implementation and management. 

The Reactive Non-uniform Cache Access (R-NUCA) machine~\cite{reactivenuca} also seeks to alleviate data movement overhead by combining different caches to form \textit{clusters}, and dynamically adjusts the cluster size based on software-defined parameters.
R-NUCA attempts to evenly distribute the data across different caches within a cluster to ensure even access latency. Our proposal achieves a similar goal for PIM systems by moving data next to the processsing elements that need it.

% The Reactive Non-uniform Cache Access (R-NUCA) machine \cite{reactivenuca} also seeks to alleviate data movement overhead by combining different caches to form \textit{clusters}, and dynamically adjusts the cluster size based on software-defined parameters. R-NUCA attempts to evenly distribute the data across different caches within a cluster, to ensure even access latency.

\section{Conclusions}
\label{sec_conclusions}

While PIM systems aim to minimize data movements between processors and memory, they often incur significant delays due to processing data in remote memory modules. Our study highlights that the majority of latency per memory request arises from data transfer and queuing delays among different memory modules (vaults). To mitigate this issue, we introduce DL-PIM, a subscription-based architecture designed to enhance PIM system efficiency by increasing local accesses for near-memory processing elements.

DL-PIM incorporates an \textit{always-subscribe} policy, which proactively moves data from remote memory modules to a reserved area in the local module upon first access. Additionally, we propose an \textit{adaptive} policy that mitigates performance degradation for some workloads caused by the extra traffic of \textit{always-subscribe}. In our evaluation with HMC, our adaptive policy demonstrates an average performance improvement of 6\% across DAMOV representative workloads, with a notable 15\% enhancement for workloads characterized by significant data reuse, caused by a 54\% reduction in average memory latency per request.

Furthermore, our adaptive policy applied to HBM achieves a 50\% reduction in average memory latency per request, translating to a 5\% improvement in overall system performance for data-heavy workloads and 3\% for all workloads. 

% However, DL-PIM had no impact on many workloads with poor data reuse. It also increased bandwidth demand by 14\% due to the extra traffic caused by subscriptions. 

%%%%%%% -- PAPER CONTENT ENDS -- %%%%%%%%

%%%%%%%%% -- BIB STYLE AND FILE -- %%%%%%%%
\bibliographystyle{IEEEtranS}
\bibliography{refs}

% Generated by IEEEtranS.bst, version: 1.13 (2008/09/30)
\begin{thebibliography}{100}
\providecommand{\url}[1]{#1}
\csname url@samestyle\endcsname
\providecommand{\newblock}{\relax}
\providecommand{\bibinfo}[2]{#2}
\providecommand{\BIBentrySTDinterwordspacing}{\spaceskip=0pt\relax}
\providecommand{\BIBentryALTinterwordstretchfactor}{4}
\providecommand{\BIBentryALTinterwordspacing}{\spaceskip=\fontdimen2\font plus
\BIBentryALTinterwordstretchfactor\fontdimen3\font minus \fontdimen4\font\relax}
\providecommand{\BIBforeignlanguage}[2]{{%
\expandafter\ifx\csname l@#1\endcsname\relax
\typeout{** WARNING: IEEEtranS.bst: No hyphenation pattern has been}%
\typeout{** loaded for the language `#1'. Using the pattern for}%
\typeout{** the default language instead.}%
\else
\language=\csname l@#1\endcsname
\fi
#2}}
\providecommand{\BIBdecl}{\relax}
\BIBdecl

\bibitem{computecaches}
S.~Aga, S.~Jeloka, A.~Subramaniyan, S.~Narayanasamy, D.~Blaauw, and R.~Das, ``{Compute Caches},'' in \emph{{2017 IEEE International Symposium on High Performance Computer Architecture (HPCA)}}, 2017, pp. 481--492.

\bibitem{scalablepimforgraphprocessing}
J.~Ahn, S.~Hong, S.~Yoo, O.~Mutlu, and K.~Choi, ``A scalable processing-in-memory accelerator for parallel graph processing,'' in \emph{2015 ACM/IEEE 42nd Annual International Symposium on Computer Architecture (ISCA)}, 2015, pp. 105--117.

\bibitem{pimenabledinsts}
\BIBentryALTinterwordspacing
J.~Ahn, S.~Yoo, O.~Mutlu, and K.~Choi, ``Pim-enabled instructions: A low-overhead, locality-aware processing-in-memory architecture,'' in \emph{Proceedings of the 42nd Annual International Symposium on Computer Architecture}, ser. ISCA '15.\hskip 1em plus 0.5em minus 0.4em\relax New York, NY, USA: Association for Computing Machinery, 2015, p. 336–348. [Online]. Available: \url{https://doi-org.proxy.lib.sfu.ca/10.1145/2749469.2750385}
\BIBentrySTDinterwordspacing

\bibitem{pimenabledinstructions}
\BIBentryALTinterwordspacing
J.~Ahn, S.~Yoo, O.~Mutlu, and K.~Choi, ``Pim-enabled instructions: A low-overhead, locality-aware processing-in-memory architecture,'' \emph{SIGARCH Comput. Archit. News}, vol.~43, no.~3S, p. 336–348, jun 2015. [Online]. Available: \url{https://doi-org.proxy.lib.sfu.ca/10.1145/2872887.2750385}
\BIBentrySTDinterwordspacing

\bibitem{akin2015data}
B.~Akin, F.~Franchetti, and J.~C. Hoe, ``Data reorganization in memory using 3d-stacked dram,'' \emph{ACM SIGARCH Computer Architecture News}, vol.~43, no.~3S, pp. 131--143, 2015.

\bibitem{angizi2019accelerating}
S.~Angizi and D.~Fan, ``Accelerating bulk bit-wise x(n)or operation in processing-in-dram platform,'' 2019.

\bibitem{pimalogic}
S.~Angizi, Z.~He, and D.~Fan, ``Pima-logic: A novel processing-in-memory architecture for highly flexible and energy-efficient logic computation,'' in \emph{2018 55th ACM/ESDA/IEEE Design Automation Conference (DAC)}, 2018, pp. 1--6.

\bibitem{cmppim}
S.~Angizi, Z.~He, A.~S. Rakin, and D.~Fan, ``Cmp-pim: An energy-efficient comparator-based processing-in-memory neural network accelerator,'' in \emph{2018 55th ACM/ESDA/IEEE Design Automation Conference (DAC)}, 2018, pp. 1--6.

\bibitem{aligns}
S.~Angizi, J.~Sun, W.~Zhang, and D.~Fan, ``Aligns: A processing-in-memory accelerator for dna short read alignment leveraging sot-mram,'' in \emph{2019 56th ACM/IEEE Design Automation Conference (DAC)}, 2019, pp. 1--6.

\bibitem{chameleon}
H.~Asghari-Moghaddam, Y.~H. Son, J.~H. Ahn, and N.~S. Kim, ``Chameleon: Versatile and practical near-dram acceleration architecture for large memory systems,'' in \emph{2016 49th Annual IEEE/ACM International Symposium on Microarchitecture (MICRO)}, 2016, pp. 1--13.

\bibitem{jafar}
\BIBentryALTinterwordspacing
A.~Augusta and S.~Idreos, ``Jafar: Near-data processing for databases,'' in \emph{Proceedings of the 2015 ACM SIGMOD International Conference on Management of Data}, ser. SIGMOD '15.\hskip 1em plus 0.5em minus 0.4em\relax New York, NY, USA: Association for Computing Machinery, 2015, p. 2069–2070. [Online]. Available: \url{https://doi-org.proxy.lib.sfu.ca/10.1145/2723372.2764942}
\BIBentrySTDinterwordspacing

\bibitem{hashjoin}
{\c{C}}.~Balkesen, J.~Teubner, G.~Alonso, and M.~T. {\"O}zsu, ``Main-memory hash joins on modern processor architectures,'' \emph{IEEE Transactions on Knowledge and Data Engineering}, vol.~27, no.~7, pp. 1754--1766, 2014.

\bibitem{revamp}
D.~Bhattacharjee, R.~Devadoss, and A.~Chattopadhyay, ``Revamp: Reram based vliw architecture for in-memory computing,'' in \emph{Design, Automation \& Test in Europe Conference \& Exhibition (DATE), 2017}, 2017, pp. 782--787.

\bibitem{googleworkload}
\BIBentryALTinterwordspacing
A.~Boroumand, S.~Ghose, Y.~Kim, R.~Ausavarungnirun, E.~Shiu, R.~Thakur, D.~Kim, A.~Kuusela, A.~Knies, P.~Ranganathan, and O.~Mutlu, ``Google workloads for consumer devices: Mitigating data movement bottlenecks,'' \emph{SIGPLAN Not.}, vol.~53, no.~2, p. 316–331, mar 2018. [Online]. Available: \url{https://doi-org.proxy.lib.sfu.ca/10.1145/3296957.3173177}
\BIBentrySTDinterwordspacing

\bibitem{conda}
A.~Boroumand, S.~Ghose, M.~Patel, H.~Hassan, B.~Lucia, R.~Ausavarungnirun, K.~Hsieh, N.~Hajinazar, K.~T. Malladi, H.~Zheng, and O.~Mutlu, ``Conda: Efficient cache coherence support for near-data accelerators,'' in \emph{2019 ACM/IEEE 46th Annual International Symposium on Computer Architecture (ISCA)}, 2019, pp. 629--642.

\bibitem{lazypim}
A.~Boroumand, S.~Ghose, M.~Patel, H.~Hassan, B.~Lucia, K.~Hsieh, K.~T. Malladi, H.~Zheng, and O.~Mutlu, ``Lazypim: An efficient cache coherence mechanism for processing-in-memory,'' \emph{IEEE Computer Architecture Letters}, vol.~16, no.~1, pp. 46--50, 2017.

\bibitem{drstrange}
F.~N. Bostancı, A.~Olgun, L.~Orosa, A.~G. Yağlıkçı, J.~S. Kim, H.~Hassan, O.~Ergin, and O.~Mutlu, ``Dr-strange: End-to-end system design for dram-based true random number generators,'' in \emph{2022 IEEE International Symposium on High-Performance Computer Architecture (HPCA)}, 2022, pp. 1141--1155.

\bibitem{lisa}
K.~K. Chang, P.~J. Nair, D.~Lee, S.~Ghose, M.~K. Qureshi, and O.~Mutlu, ``Low-cost inter-linked subarrays (lisa): Enabling fast inter-subarray data movement in dram,'' in \emph{2016 IEEE International Symposium on High Performance Computer Architecture (HPCA)}, 2016, pp. 568--580.

\bibitem{rodinia}
S.~Che, M.~Boyer, J.~Meng, D.~Tarjan, J.~W. Sheaffer, S.-H. Lee, and K.~Skadron, ``Rodinia: A benchmark suite for heterogeneous computing,'' in \emph{2009 IEEE International Symposium on Workload Characterization (IISWC)}, 2009, pp. 44--54.

\bibitem{prime}
P.~Chi, S.~Li, C.~Xu, T.~Zhang, J.~Zhao, Y.~Liu, Y.~Wang, and Y.~Xie, ``Prime: A novel processing-in-memory architecture for neural network computation in reram-based main memory,'' in \emph{2016 ACM/IEEE 43rd Annual International Symposium on Computer Architecture (ISCA)}, 2016, pp. 27--39.

\bibitem{candydramcaches}
C.~Chou, A.~Jaleel, and M.~K. Qureshi, ``Candy: Enabling coherent dram caches for multi-node systems,'' in \emph{2016 49th Annual IEEE/ACM International Symposium on Microarchitecture (MICRO)}, 2016, pp. 1--13.

\bibitem{bigdata-memory}
R.~Clapp, M.~Dimitrov, K.~Kumar, V.~Viswanathan, and T.~Willhalm, ``Quantifying the performance impact of memory latency and bandwidth for big data workloads,'' in \emph{2015 IEEE International Symposium on Workload Characterization}, 2015, pp. 213--224.

\bibitem{coma}
F.~Dahlgren and J.~Torrellas, ``Cache-only memory architectures,'' \emph{Computer}, vol.~32, no.~6, pp. 72--79, 1999.

\bibitem{graphh}
G.~Dai, T.~Huang, Y.~Chi, J.~Zhao, G.~Sun, Y.~Liu, Y.~Wang, Y.~Xie, and H.~Yang, ``Graphh: A processing-in-memory architecture for large-scale graph processing,'' \emph{IEEE Transactions on Computer-Aided Design of Integrated Circuits and Systems}, vol.~38, no.~4, pp. 640--653, 2019.

\bibitem{upmemtruepimaccelerator}
F.~Devaux, ``The true processing in memory accelerator,'' in \emph{2019 IEEE Hot Chips 31 Symposium (HCS)}, 2019, pp. 1--24.

\bibitem{neuralcache}
C.~Eckert, X.~Wang, J.~Wang, A.~Subramaniyan, R.~Iyer, D.~Sylvester, D.~Blaaauw, and R.~Das, ``{Neural Cache: Bit-Serial In-Cache Acceleration of Deep Neural Networks},'' in \emph{{2018 ACM/IEEE 45th Annual International Symposium on Computer Architecture (ISCA)}}, 2018, pp. 383--396.

\bibitem{reactivenuma}
B.~Falsafi and D.~A. Wood, ``Reactive numa: A design for unifying s-coma and cc-numa,'' \emph{SIGARCH Comput. Archit. News}, vol.~25, no.~2, p. 229–240, May 1997.

\bibitem{nda}
A.~Farmahini-Farahani, J.~H. Ahn, K.~Morrow, and N.~S. Kim, ``Nda: Near-dram acceleration architecture leveraging commodity dram devices and standard memory modules,'' in \emph{2015 IEEE 21st International Symposium on High Performance Computer Architecture (HPCA)}, 2015, pp. 283--295.

\bibitem{dualitycache}
D.~Fujiki, S.~Mahlke, and R.~Das, ``{Duality Cache for Data Parallel Acceleration},'' in \emph{{Proceedings of the 46th International Symposium on Computer Architecture}}.\hskip 1em plus 0.5em minus 0.4em\relax New York, NY, USA: Association for Computing Machinery, 2019, pp. 397--410.

\bibitem{plim}
P.-E. Gaillardon, L.~Amarú, A.~Siemon, E.~Linn, R.~Waser, A.~Chattopadhyay, and G.~De~Micheli, ``The programmable logic-in-memory (plim) computer,'' in \emph{2016 Design, Automation \& Test in Europe Conference \& Exhibition (DATE)}, 2016, pp. 427--432.

\bibitem{practicalneardataprocessing}
M.~Gao, G.~Ayers, and C.~Kozyrakis, ``Practical near-data processing for in-memory analytics frameworks,'' in \emph{2015 International Conference on Parallel Architecture and Compilation (PACT)}, 2015, pp. 113--124.

\bibitem{hrl}
M.~Gao and C.~Kozyrakis, ``Hrl: Efficient and flexible reconfigurable logic for near-data processing,'' in \emph{2016 IEEE International Symposium on High Performance Computer Architecture (HPCA)}, 2016, pp. 126--137.

\bibitem{tetrisnn}
M.~Gao, J.~Pu, X.~Yang, M.~Horowitz, and C.~Kozyrakis, ``Tetris: Scalable and efficient neural network acceleration with 3d memory,'' in \emph{Proceedings of the Twenty-Second International Conference on Architectural Support for Programming Languages and Operating Systems}.\hskip 1em plus 0.5em minus 0.4em\relax New York, NY, USA: Association for Computing Machinery, 2017, pp. 751--764.

\bibitem{pimworkloaddriven}
S.~Ghose, A.~Boroumand, J.~S. Kim, J.~Gómez-Luna, and O.~Mutlu, ``{Processing-in-memory: A workload-driven perspective},'' \emph{{IBM Journal of Research and Development}}, vol.~63, no.~6, pp. 3:1--3:19, 2019.

\bibitem{biscuit}
B.~Gu, A.~S. Yoon, D.-H. Bae, I.~Jo, J.~Lee, J.~Yoon, J.-U. Kang, M.~Kwon, C.~Yoon, S.~Cho, J.~Jeong, and D.~Chang, ``Biscuit: A framework for near-data processing of big data workloads,'' in \emph{2016 ACM/IEEE 43rd Annual International Symposium on Computer Architecture (ISCA)}, 2016, pp. 153--165.

\bibitem{bimodaldramcache}
N.~Gulur, M.~Mehendale, R.~Manikantan, and R.~Govindarajan, ``Bi-modal dram cache: Improving hit rate, hit latency and bandwidth,'' in \emph{2014 47th Annual IEEE/ACM International Symposium on Microarchitecture}, 2014, pp. 38--50.

\bibitem{guo20143d}
Q.~Guo, N.~Alachiotis, B.~Akin, F.~Sadi, G.~Xu, T.-M. Low, L.~Pileggi, J.~C. Hoe, and F.~Franchetti, ``3d-stacked memory-side acceleration: Accelerator and system design,'' 2014.

\bibitem{chai}
J.~Gómez-Luna, I.~E. Hajj, L.-W. Chang, V.~García-Floreszx, S.~G. de~Gonzalo, T.~B. Jablin, A.~J. Peña, and W.-m. Hwu, ``Chai: Collaborative heterogeneous applications for integrated-architectures,'' in \emph{2017 IEEE International Symposium on Performance Analysis of Systems and Software (ISPASS)}, 2017, pp. 43--54.

\bibitem{simdram}
N.~Hajinazar, G.~F. Oliveira, S.~Gregorio, J.~a.~D. Ferreira, N.~M. Ghiasi, M.~Patel, M.~Alser, S.~Ghose, J.~G\'{o}mez-Luna, and O.~Mutlu, ``Simdram: A framework for bit-serial simd processing using dram,'' in \emph{Proceedings of the 26th ACM International Conference on Architectural Support for Programming Languages and Operating Systems}, ser. ASPLOS '21.\hskip 1em plus 0.5em minus 0.4em\relax New York, NY, USA: Association for Computing Machinery, 2021, p. 329–345.

\bibitem{memristorforcomputing}
S.~Hamdioui, S.~Kvatinsky, G.~Cauwenberghs, L.~Xie, N.~Wald, S.~Joshi, H.~M. Elsayed, H.~Corporaal, and K.~Bertels, ``Memristor for computing: Myth or reality?'' in \emph{Design, Automation \& Test in Europe Conference \& Exhibition (DATE), 2017}, 2017, pp. 722--731.

\bibitem{memristorbasedcomputation}
S.~Hamdioui, L.~Xie, H.~A. Du~Nguyen, M.~Taouil, K.~Bertels, H.~Corporaal, H.~Jiao, F.~Catthoor, D.~Wouters, L.~Eike, and J.~van Lunteren, ``Memristor based computation-in-memory architecture for data-intensive applications,'' in \emph{2015 Design, Automation \& Test in Europe Conference \& Exhibition (DATE)}, 2015, pp. 1718--1725.

\bibitem{reactivenuca}
N.~Hardavellas, M.~Ferdman, B.~Falsafi, and A.~Ailamaki, ``Reactive nuca: Near-optimal block placement and replication in distributed caches,'' in \emph{Proceedings of the 36th Annual International Symposium on Computer Architecture}.\hskip 1em plus 0.5em minus 0.4em\relax New York, NY, USA: Association for Computing Machinery, 2009, pp. 184--195.

\bibitem{enhancedmemorycontroller}
M.~Hashemi, Khubaib, E.~Ebrahimi, O.~Mutlu, and Y.~N. Patt, ``Accelerating dependent cache misses with an enhanced memory controller,'' in \emph{2016 ACM/IEEE 43rd Annual International Symposium on Computer Architecture (ISCA)}, 2016, pp. 444--455.

\bibitem{continuousrenahead}
M.~Hashemi, O.~Mutlu, and Y.~N. Patt, ``Continuous runahead: Transparent hardware acceleration for memory intensive workloads,'' in \emph{2016 49th Annual IEEE/ACM International Symposium on Microarchitecture (MICRO)}, 2016, pp. 1--12.

\bibitem{impica}
K.~Hsieh, S.~Khan, N.~Vijaykumar, K.~K. Chang, A.~Boroumand, S.~Ghose, and O.~Mutlu, ``Accelerating pointer chasing in 3d-stacked memory: Challenges, mechanisms, evaluation,'' in \emph{2016 IEEE 34th International Conference on Computer Design (ICCD)}, 2016, pp. 25--32.

\bibitem{atcachesramtagcache}
C.-C. Huang and V.~Nagarajan, ``Atcache: Reducing dram cache latency via a small sram tag cache,'' in \emph{2014 23rd International Conference on Parallel Architecture and Compilation Techniques (PACT)}, 2014, pp. 51--60.

\bibitem{transpimlib}
M.~Item, G.~F. Oliveira, J.~Gómez-Luna, M.~Sadrosadati, Y.~Guo, and O.~Mutlu, ``Transpimlib: Efficient transcendental functions for processing-in-memory systems,'' in \emph{2023 IEEE International Symposium on Performance Analysis of Systems and Software (ISPASS)}, 2023, pp. 235--247.

\bibitem{jun2017hbm}
H.~Jun, J.~Cho, K.~Lee, H.-Y. Son, K.~Kim, H.~Jin, and K.~Kim, ``Hbm (high bandwidth memory) dram technology and architecture,'' in \emph{2017 IEEE International Memory Workshop (IMW)}.\hskip 1em plus 0.5em minus 0.4em\relax IEEE, 2017, pp. 1--4.

\bibitem{energyefficientvlsi}
M.~Kang, M.-S. Keel, N.~R. Shanbhag, S.~Eilert, and K.~Curewitz, ``An energy-efficient vlsi architecture for pattern recognition via deep embedding of computation in sram,'' in \emph{2014 IEEE International Conference on Acoustics, Speech and Signal Processing (ICASSP)}, 2014, pp. 8326--8330.

\bibitem{axdimm}
L.~Ke, X.~Zhang, J.~So, J.-G. Lee, S.-H. Kang, S.~Lee, S.~Han, Y.~Cho, J.~H. Kim, Y.~Kwon, K.~Kim, J.~Jung, I.~Yun, S.~J. Park, H.~Park, J.~Song, J.~Cho, K.~Sohn, N.~S. Kim, and H.-H.~S. Lee, ``Near-memory processing in action: Accelerating personalized recommendation with axdimm,'' \emph{IEEE Micro}, vol.~42, no.~1, pp. 116--127, 2022.

\bibitem{neurocube}
D.~Kim, J.~Kung, S.~Chai, S.~Yalamanchili, and S.~Mukhopadhyay, ``Neurocube: A programmable digital neuromorphic architecture with high-density 3d memory,'' in \emph{2016 ACM/IEEE 43rd Annual International Symposium on Computer Architecture (ISCA)}, 2016, pp. 380--392.

\bibitem{kim2018grim}
J.~S. Kim, D.~Senol~Cali, H.~Xin, D.~Lee, S.~Ghose, M.~Alser, H.~Hassan, O.~Ergin, C.~Alkan, and O.~Mutlu, ``Grim-filter: Fast seed location filtering in dna read mapping using processing-in-memory technologies,'' \emph{BMC genomics}, vol.~19, no.~2, pp. 23--40, 2018.

\bibitem{ramulator}
Y.~Kim, W.~Yang, and O.~Mutlu, ``Ramulator: A fast and extensible dram simulator,'' \emph{IEEE Computer Architecture Letters}, vol.~15, no.~1, pp. 45--49, 2016.

\bibitem{magic}
S.~Kvatinsky, D.~Belousov, S.~Liman, G.~Satat, N.~Wald, E.~G. Friedman, A.~Kolodny, and U.~C. Weiser, ``Magic—memristor-aided logic,'' \emph{IEEE Transactions on Circuits and Systems II: Express Briefs}, vol.~61, no.~11, pp. 895--899, 2014.

\bibitem{imply}
S.~Kvatinsky, A.~Kolodny, U.~C. Weiser, and E.~G. Friedman, ``Memristor-based imply logic design procedure,'' in \emph{2011 IEEE 29th International Conference on Computer Design (ICCD)}, 2011, pp. 142--147.

\bibitem{implyprinciplesandmethodologies}
S.~Kvatinsky, G.~Satat, N.~Wald, E.~G. Friedman, A.~Kolodny, and U.~C. Weiser, ``Memristor-based material implication (imply) logic: Design principles and methodologies,'' \emph{IEEE Transactions on Very Large Scale Integration (VLSI) Systems}, vol.~22, no.~10, pp. 2054--2066, 2014.

\bibitem{a20nm}
Y.-C. Kwon, S.~H. Lee, J.~Lee, S.-H. Kwon, J.~M. Ryu, J.-P. Son, O.~Seongil, H.-S. Yu, H.~Lee, S.~Y. Kim, Y.~Cho, J.~G. Kim, J.~Choi, H.-S. Shin, J.~Kim, B.~Phuah, H.~Kim, M.~J. Song, A.~Choi, D.~Kim, S.~Kim, E.-B. Kim, D.~Wang, S.~Kang, Y.~Ro, S.~Seo, J.~Song, J.~Youn, K.~Sohn, and N.~S. Kim, ``A 20nm 6gb function-in-memory dram, based on hbm2 with a 1.2tflops programmable computing unit using bank-level parallelism, for machine learning applications,'' in \emph{2021 IEEE International Solid- State Circuits Conference (ISSCC)}, vol.~64, 2021, pp. 350--352.

\bibitem{thesgiorigin}
J.~Laudon and D.~Lenoski, ``The sgi origin: A ccnuma highly scalable server,'' \emph{SIGARCH Comput. Archit. News}, vol.~25, no.~2, pp. 241--251, May 1997.

\bibitem{lee20141}
D.~U. Lee, K.~W. Kim, K.~W. Kim, K.~S. Lee, S.~J. Byeon, J.~H. Kim, J.~H. Cho, J.~Lee, and J.~H. Chun, ``A 1.2 v 8 gb 8-channel 128 gb/s high-bandwidth memory (hbm) stacked dram with effective i/o test circuits,'' \emph{IEEE Journal of Solid-State Circuits}, vol.~50, no.~1, pp. 191--203, 2014.

\bibitem{bssync}
J.~H. Lee, J.~Sim, and H.~Kim, ``Bssync: Processing near memory for machine learning workloads with bounded staleness consistency models,'' in \emph{2015 International Conference on Parallel Architecture and Compilation (PACT)}, 2015, pp. 241--252.

\bibitem{industrialproduct}
S.~Lee, S.-h. Kang, J.~Lee, H.~Kim, E.~Lee, S.~Seo, H.~Yoon, S.~Lee, K.~Lim, H.~Shin, J.~Kim, O.~Seongil, A.~Iyer, D.~Wang, K.~Sohn, and N.~S. Kim, ``Hardware architecture and software stack for pim based on commercial dram technology : Industrial product,'' in \emph{2021 ACM/IEEE 48th Annual International Symposium on Computer Architecture (ISCA)}, 2021, pp. 43--56.

\bibitem{fataglessdramcache}
Y.~Lee, J.~Kim, H.~Jang, H.~Yang, J.~Kim, J.~Jeong, and J.~W. Lee, ``A fully associative, tagless dram cache,'' in \emph{2015 ACM/IEEE 42nd Annual International Symposium on Computer Architecture (ISCA)}, 2015, pp. 211--222.

\bibitem{dashprotocol}
D.~Lenoski, J.~Laudon, K.~Gharachorloo, A.~Gupta, and J.~Hennessy, ``The directory-based cache coherence protocol for the dash multiprocessor,'' \emph{SIGARCH Comput. Archit. News}, vol.~18, no. 2SI, pp. 148--159, may 1990.

\bibitem{levy_logic_2014}
\BIBentryALTinterwordspacing
Y.~Levy, J.~Bruck, Y.~Cassuto, E.~G. Friedman, A.~Kolodny, E.~Yaakobi, and S.~Kvatinsky, ``Logic operations in memory using a memristive akers array,'' \emph{Microelectronics Journal}, vol.~45, no.~11, pp. 1429--1437, 2014. [Online]. Available: \url{https://www.sciencedirect.com/science/article/pii/S0026269214002055}
\BIBentrySTDinterwordspacing

\bibitem{drisa}
S.~Li, D.~Niu, K.~T. Malladi, H.~Zheng, B.~Brennan, and Y.~Xie, ``Drisa: A dram-based reconfigurable in-situ accelerator,'' in \emph{2017 50th Annual IEEE/ACM International Symposium on Microarchitecture (MICRO)}, 2017, pp. 288--301.

\bibitem{pinatubo}
S.~Li, C.~Xu, Q.~Zou, J.~Zhao, Y.~Lu, and Y.~Xie, ``Pinatubo: A processing-in-memory architecture for bulk bitwise operations in emerging non-volatile memories,'' in \emph{2016 53nd ACM/EDAC/IEEE Design Automation Conference (DAC)}, 2016, pp. 1--6.

\bibitem{effectivenessof3dstackedcaches}
G.~H. Loh, ``Extending the effectiveness of 3d-stacked dram caches with an adaptive multi-queue policy,'' in \emph{2009 42nd Annual IEEE/ACM International Symposium on Microarchitecture (MICRO)}, 2009, pp. 201--212.

\bibitem{enableconventionalblocksizes}
G.~H. Loh and M.~D. Hill, ``Efficiently enabling conventional block sizes for very large die-stacked dram caches,'' in \emph{2011 44th Annual IEEE/ACM International Symposium on Microarchitecture (MICRO)}, 2011, pp. 454--464.

\bibitem{genstore}
\BIBentryALTinterwordspacing
N.~Mansouri~Ghiasi, J.~Park, H.~Mustafa, J.~Kim, A.~Olgun, A.~Gollwitzer, D.~Senol~Cali, C.~Firtina, H.~Mao, N.~Almadhoun~Alserr, R.~Ausavarungnirun, N.~Vijaykumar, M.~Alser, and O.~Mutlu, ``Genstore: A high-performance in-storage processing system for genome sequence analysis,'' in \emph{Proceedings of the 27th ACM International Conference on Architectural Support for Programming Languages and Operating Systems}, ser. ASPLOS '22.\hskip 1em plus 0.5em minus 0.4em\relax New York, NY, USA: Association for Computing Machinery, 2022, p. 635–654. [Online]. Available: \url{https://doi-org.proxy.lib.sfu.ca/10.1145/3503222.3507702}
\BIBentrySTDinterwordspacing

\bibitem{mccalpin1995memory}
J.~D. McCalpin \emph{et~al.}, ``Memory bandwidth and machine balance in current high performance computers,'' \emph{IEEE computer society technical committee on computer architecture (TCCA) newsletter}, vol.~2, no. 19-25, 1995.

\bibitem{reflectiononmemwall}
S.~A. McKee, ``Reflections on the memory wall,'' in \emph{Proceedings of the 1st Conference on Computing Frontiers}, ser. CF '04.\hskip 1em plus 0.5em minus 0.4em\relax New York, NY, USA: Association for Computing Machinery, 2004, p. 162.

\bibitem{micron}
\BIBentryALTinterwordspacing
{Micron Technology, Inc.}, ``{Hybrid Memory Cube – HMC Gen2}.'' [Online]. Available: \url{https://www.micron.com/-/media/client/global/documents/products/data-sheet/hmc/gen2/hmc_gen2.pdf}
\BIBentrySTDinterwordspacing

\bibitem{muthukrishnan2021gps}
H.~Muthukrishnan, D.~Lustig, D.~Nellans, and T.~Wenisch, ``Gps: A global publish-subscribe model for multi-gpu memory management,'' in \emph{MICRO-54: 54th Annual IEEE/ACM International Symposium on Microarchitecture}, 2021, pp. 46--58.

\bibitem{mutlu2021intelligent}
O.~Mutlu, ``Intelligent architectures for intelligent computing systems,'' in \emph{2021 Design, Automation \& Test in Europe Conference \& Exhibition (DATE)}.\hskip 1em plus 0.5em minus 0.4em\relax IEEE, 2021, pp. 318--323.

\bibitem{oh202022}
C.-S. Oh, K.~C. Chun, Y.-Y. Byun, Y.-K. Kim, S.-Y. Kim, Y.~Ryu, J.~Park, S.~Kim, S.~Cha, D.~Shin \emph{et~al.}, ``22.1 a 1.1 v 16gb 640gb/s hbm2e dram with a data-bus window-extension technique and a synergetic on-die ecc scheme,'' in \emph{2020 IEEE International Solid-State Circuits Conference-(ISSCC)}.\hskip 1em plus 0.5em minus 0.4em\relax IEEE, 2020, pp. 330--332.

\bibitem{oliveira2021damov}
G.~F. Oliveira, J.~G{\'o}mez-Luna, L.~Orosa, S.~Ghose, N.~Vijaykumar, I.~Fernandez, M.~Sadrosadati, and O.~Mutlu, ``{DAMOV: A new methodology and benchmark suite for evaluating data movement bottlenecks},'' \emph{IEEE Access}, vol.~9, pp. 134\,457--134\,502, 2021.

\bibitem{flashcosmos}
J.~Park, R.~Azizi, G.~F. Oliveira, M.~Sadrosadati, R.~Nadig, D.~Novo, J.~Gómez-Luna, M.~Kim, and O.~Mutlu, ``Flash-cosmos: In-flash bulk bitwise operations using inherent computation capability of nand flash memory,'' in \emph{2022 55th IEEE/ACM International Symposium on Microarchitecture (MICRO)}, 2022, pp. 937--955.

\bibitem{polybench}
L.-N. Pouchet, ``Polybench: The polyhedral benchmark suite,'' \url{https://www.cs.colostate.edu/~pouchet/software/polybench/}, 2011--2012.

\bibitem{fundamentallattradeoff}
M.~K. Qureshi and G.~H. Loh, ``Fundamental latency trade-off in architecting dram caches: Outperforming impractical sram-tags with a simple and practical design,'' in \emph{2012 45th Annual IEEE/ACM International Symposium on Microarchitecture}, 2012, pp. 235--246.

\bibitem{mlpawarecacherep}
M.~K. Qureshi, D.~N. Lynch, O.~Mutlu, and Y.~N. Patt, ``A case for mlp-aware cache replacement,'' in \emph{Proceedings of the 33rd Annual International Symposium on Computer Architecture}.\hskip 1em plus 0.5em minus 0.4em\relax USA: IEEE Computer Society, 2006, pp. 167--178.

\bibitem{darknet13}
J.~Redmon, ``Darknet: Open source neural networks in c,'' \url{http://pjreddie.com/darknet/}, 2013--2016.

\bibitem{zsim}
D.~Sanchez and C.~Kozyrakis, ``Zsim: Fast and accurate microarchitectural simulation of thousand-core systems,'' in \emph{Proceedings of the 40th Annual International Symposium on Computer Architecture}.\hskip 1em plus 0.5em minus 0.4em\relax New York, NY, USA: Association for Computing Machinery, 2013, pp. 475--486.

\bibitem{missingthememwall}
A.~Saulsbury, F.~Pong, and A.~Nowatzyk, ``Missing the memory wall: The case for processor/memory integration,'' \emph{SIGARCH Comput. Archit. News}, vol.~24, no.~2, pp. 90--101, May 1996.

\bibitem{fastbulkbitwiseandandor}
V.~Seshadri, K.~Hsieh, A.~Boroum, D.~Lee, M.~A. Kozuch, O.~Mutlu, P.~B. Gibbons, and T.~C. Mowry, ``Fast bulk bitwise and and or in dram,'' \emph{IEEE Computer Architecture Letters}, vol.~14, no.~2, pp. 127--131, 2015.

\bibitem{rowclone}
V.~Seshadri, Y.~Kim, C.~Fallin, D.~Lee, R.~Ausavarungnirun, G.~Pekhimenko, Y.~Luo, O.~Mutlu, P.~B. Gibbons, M.~A. Kozuch, and T.~C. Mowry, ``Rowclone: Fast and energy-efficient in-dram bulk data copy and initialization,'' in \emph{2013 46th Annual IEEE/ACM International Symposium on Microarchitecture (MICRO)}, 2013, pp. 185--197.

\bibitem{seshadri2016buddyram}
V.~Seshadri, D.~Lee, T.~Mullins, H.~Hassan, A.~Boroumand, J.~Kim, M.~A. Kozuch, O.~Mutlu, P.~B. Gibbons, and T.~C. Mowry, ``Buddy-ram: Improving the performance and efficiency of bulk bitwise operations using dram,'' 2016.

\bibitem{ambit}
V.~Seshadri, D.~Lee, T.~Mullins, H.~Hassan, A.~Boroumand, J.~Kim, M.~A. Kozuch, O.~Mutlu, P.~B. Gibbons, and T.~C. Mowry, ``Ambit: In-memory accelerator for bulk bitwise operations using commodity dram technology,'' in \emph{2017 50th Annual IEEE/ACM International Symposium on Microarchitecture (MICRO)}, 2017, pp. 273--287.

\bibitem{gatherscatterdram}
V.~Seshadri, T.~Mullins, A.~Boroumand, O.~Mutlu, P.~B. Gibbons, M.~A. Kozuch, and T.~C. Mowry, ``Gather-scatter dram: In-dram address translation to improve the spatial locality of non-unit strided accesses,'' in \emph{2015 48th Annual IEEE/ACM International Symposium on Microarchitecture (MICRO)}, 2015, pp. 267--280.

\bibitem{seshadri2016processing}
V.~Seshadri and O.~Mutlu, ``The processing using memory paradigm:in-dram bulk copy, initialization, bitwise and and or,'' 2016.

\bibitem{isaac}
A.~Shafiee, A.~Nag, N.~Muralimanohar, R.~Balasubramonian, J.~P. Strachan, M.~Hu, R.~S. Williams, and V.~Srikumar, ``Isaac: A convolutional neural network accelerator with in-situ analog arithmetic in crossbars,'' in \emph{2016 ACM/IEEE 43rd Annual International Symposium on Computer Architecture (ISCA)}, 2016, pp. 14--26.

\bibitem{ligra}
J.~Shun and G.~E. Blelloch, ``Ligra: A lightweight graph processing framework for shared memory,'' \emph{SIGPLAN Not.}, vol.~48, no.~8, pp. 135--146, Feb 2013.

\bibitem{mostlycleandramcache}
J.~Sim, G.~H. Loh, H.~Kim, M.~OConnor, and M.~Thottethodi, ``A mostly-clean dram cache for effective hit speculation and self-balancing dispatch,'' in \emph{2012 45th Annual IEEE/ACM International Symposium on Microarchitecture}, 2012, pp. 247--257.

\bibitem{fpgabasednmp}
G.~Singh, M.~Alser, D.~S. Cali, D.~Diamantopoulos, J.~Gómez-Luna, H.~Corporaal, and O.~Mutlu, ``Fpga-based near-memory acceleration of modern data-intensive applications,'' \emph{IEEE Micro}, vol.~41, no.~4, pp. 39--48, 2021.

\bibitem{napel}
G.~Singh, J.~Gómez-Luna, G.~Mariani, G.~F. Oliveira, S.~Corda, S.~Stuijk, O.~Mutlu, and H.~Corporaal, ``Napel: Near-memory computing application performance prediction via ensemble learning,'' in \emph{2019 56th ACM/IEEE Design Automation Conference (DAC)}, 2019, pp. 1--6.

\bibitem{sohn20161}
K.~Sohn, W.-J. Yun, R.~Oh, C.-S. Oh, S.-Y. Seo, M.-S. Park, D.-H. Shin, W.-C. Jung, S.-H. Shin, J.-M. Ryu \emph{et~al.}, ``A 1.2 v 20 nm 307 gb/s hbm dram with at-speed wafer-level io test scheme and adaptive refresh considering temperature distribution,'' \emph{IEEE Journal of Solid-State Circuits}, vol.~52, no.~1, pp. 250--260, 2016.

\bibitem{comparativeofccnumaandcoma}
P.~Stenstrom, T.~Joe, and A.~Gupta, ``Comparative performance evaluation of cache-coherent numa and coma architectures,'' in \emph{[1992] Proceedings the 19th Annual International Symposium on Computer Architecture}, 1992, pp. 80--91.

\bibitem{tian2023abndp}
B.~Tian, Q.~Chen, and M.~Gao, ``Abndp: Co-optimizing data access and load balance in near-data processing,'' in \emph{Proceedings of the 28th ACM International Conference on Architectural Support for Programming Languages and Operating Systems, Volume 3}, 2023, pp. 3--17.

\bibitem{splash2}
S.~Woo, M.~Ohara, E.~Torrie, J.~Singh, and A.~Gupta, ``The splash-2 programs: characterization and methodological considerations,'' in \emph{Proceedings 22nd Annual International Symposium on Computer Architecture}, 1995, pp. 24--36.

\bibitem{fastboolean}
L.~Xie, H.~A.~D. Nguyen, M.~Taouil, S.~Hamdioui, and K.~Bertels, ``Fast boolean logic mapped on memristor crossbar,'' in \emph{2015 33rd IEEE International Conference on Computer Design (ICCD)}, 2015, pp. 335--342.

\bibitem{spacea}
X.~Xie, Z.~Liang, P.~Gu, A.~Basak, L.~Deng, L.~Liang, X.~Hu, and Y.~Xie, ``Spacea: Sparse matrix vector multiplication on processing-in-memory accelerator,'' in \emph{2021 IEEE International Symposium on High-Performance Computer Architecture (HPCA)}, 2021, pp. 570--583.

\bibitem{phoenix}
R.~M. Yoo, A.~Romano, and C.~Kozyrakis, ``Phoenix rebirth: Scalable mapreduce on a large-scale shared-memory system,'' in \emph{2009 IEEE International Symposium on Workload Characterization (IISWC)}, 2009, pp. 198--207.

\bibitem{accord}
V.~Young, C.~Chou, A.~Jaleel, and M.~Qureshi, ``Accord: Enabling associativity for gigascale dram caches by coordinating way-install and way-prediction,'' in \emph{2018 ACM/IEEE 45th Annual International Symposium on Computer Architecture (ISCA)}, 2018, pp. 328--339.

\bibitem{memristivedevices}
J.~Yu, H.~A.~D. Nguyen, L.~Xie, M.~Taouil, and S.~Hamdioui, ``Memristive devices for computation-in-memory,'' in \emph{2018 Design, Automation \& Test in Europe Conference \& Exhibition (DATE)}, 2018, pp. 1646--1651.

\bibitem{banshee}
X.~Yu, C.~J. Hughes, N.~Satish, O.~Mutlu, and S.~Devadas, ``Banshee: Bandwidth-efficient dram caching via software/hardware cooperation,'' in \emph{2017 50th Annual IEEE/ACM International Symposium on Microarchitecture (MICRO)}, 2017, pp. 1--14.

\bibitem{memnet}
J.~Zhan, I.~Akgun, J.~Zhao, A.~Davis, P.~Faraboschi, Y.~Wang, and Y.~Xie, ``A unified memory network architecture for in-memory computing in commodity servers,'' in \emph{2016 49th Annual IEEE/ACM International Symposium on Microarchitecture (MICRO)}, 2016, pp. 1--14.

\bibitem{graphp}
M.~Zhang, Y.~Zhuo, C.~Wang, M.~Gao, Y.~Wu, K.~Chen, C.~Kozyrakis, and X.~Qian, ``Graphp: Reducing communication for pim-based graph processing with efficient data partition,'' in \emph{2018 IEEE International Symposium on High Performance Computer Architecture (HPCA)}, 2018, pp. 544--557.

\bibitem{graphq}
Y.~Zhuo, C.~Wang, M.~Zhang, R.~Wang, D.~Niu, Y.~Wang, and X.~Qian, ``Graphq: Scalable pim-based graph processing,'' in \emph{Proceedings of the 52nd Annual IEEE/ACM International Symposium on Microarchitecture}.\hskip 1em plus 0.5em minus 0.4em\relax New York, NY, USA: Association for Computing Machinery, 2019, pp. 712--725.

\end{thebibliography}
%%%%%%%%%%%%%%%%%%%%%%%%%%%%%%%%%%%%

\end{document}